\documentclass[10pt]{iopart}
\usepackage{amsmath}
\usepackage{graphicx}
\usepackage{subfigure}
\usepackage{epsfig}

\begin{document}

\title[Conformal mappings for solving ODEs]{Conformal mappings versus other
power series methods for solving ordinary differential equations:
illustration on anharmonic oscillators.}
\author{C.\ Bervillier \\
%EndAName
Laboratoire de Math\'{e}matiques et Physique Th\'{e}orique,\\
UMR 6083 (CNRS),\\
F\'{e}d\'{e}ration Denis Poisson,\\
Universit\'{e} Fran\c{c}ois Rabelais,\\
Parc de Grandmont, 37200 Tours, France.}

\begin{abstract}
The simplicity and the efficiency of a quasi-analytical method for solving
nonlinear\ ordinary differential equations (ODE), is illustrated on the
study of anharmonic oscillators (AO) with a potential $V\left( x\right)
=\beta x^{2}+x^{2m}$ ($m>0$). The method [Nucl. Phys. B801, 296 (2008)],
applies a priori to any ODE with two-point boundaries (one being located at
infinity), the solution of which has singularities in the complex plane of
the independent variable $x$. A conformal mapping of a suitably chosen
angular sector of the complex plane of $x$ upon the unit disc centered at
the origin makes convergent the transformed Taylor series of the generic
solution so that the boundary condition at infinity can be easily imposed.
In principle, this constraint, when applied on the logarithmic-derivative of
the wave function, determines the eigenvalues to an arbitrary level of
accuracy. In practice, for $\beta \geq 0$ or slightly negative, the accuracy
of the results obtained is astonishingly large with regards to the modest
computing power used. It is explained why the efficiency of the method
decreases as $\beta $ is more and more negative. Various aspects of the
method and comparisons with some seemingly similar methods, based also on
expressing the solution as a Taylor series, are shortly reviewed,\ presented
and discussed.
\end{abstract}

\pacs{02.60.-x, 02.60.Lj, 02.30.Hq, 02.30.Mv}

\submitto{\JPA}

\ead{claude.bervillier@lmpt.univ-tours.fr}

\section{Introduction}

The accurate determination of the spectrum of the anharmonic oscillator (AO)
is an old problem which has attracted much interest (for recent reviews see 
\cite{6344}). This is certainly due to the asymptotic character of its
perturbative expansion \cite{6361} (a property shared with field theory),
but also because it is intrinsically hard to solve accurately for sizeable
values of the coupling parameter. Of course, the eigenvalues may be
numerically determined using shooting or relaxation (or similar) methods,
and extremely accurate results may be obtained that way with the recourse to
large computers and sophisticated algorithms (e. g., see \cite{6428}).
Beyond the fact that a purely numerical approach does not allow to learn
much on the general properties of a solution, it often requires\ some
specific ability and experience. Now in modern physics we are more and more
confronted with the problem of solving differential equations when we are
not experts in numerical treatments. It is thus interesting to develop
simple and efficient analytical (or quasi-analytical) methods which may be
implemented using symbolic calculation softwares and modest computation
powers \cite{computer}.

It is not by accident that the mapping method \cite{6319} discussed in the
present paper has been elaborated in the context of the derivative expansion
of the exact renormalization group equation (for reviews and recent
pedagogical introductions see \cite{4595}). In that field, one is faced with
coupled nonlinear differential equations and quasi-analytical methods have
soon been introduced \cite{3478,3642} to solve them because they are much
more user-friendly than purely numerical methods. Unfortunately they were
too simplistic and do not converge \cite{3553,6201}.

More sophisticated methods have recently been proposed \cite{6110,6319}%
. They are all based on a Taylor series in the independent variable $x$ of
the solution of the ODE. Methods of that kind had previously been developped
in different contexts \cite{6201,6394}. Among all these methods, the mapping
method seems to be particularly efficient \cite{6319}. Considering first the
logarithmic derivative of the wave function (which softens the singularity
at infinity), the mapping method relies upon a conformal mapping onto the
unit disc of an angular sector of the complex plane of the independent
variable \cite{6319}. The angular sector is chosen so that the solution
looked for is analytic in its interior or assumed to be so. The mapping
compactifies the originally infinite integration domain whereas the
convergent character of the series expansion is transferred to the whole
disc of unit radius in the new variable. The condition at infinity may thus
be easily imposed. The object of the present paper is to illustrate the
simplicity and the efficiency (and also the limitations) of the mapping
method with the computation of the spectrum of the AO in different
configurations.

The paper is organized as follows. In section \ref{ODEAO}, the AO is
introduced together with a change of function (a logarithmic derivative)
which transforms the originally linear ODE into a nonlinear ODE. In doing so
the\ essential singularity at infinity of the wave functions is softened
but, at the same time, the new functions are no longer analytic in the
complex $x$-plane. The common characteristics of the Taylor-series-based
methods are then presented. (By the way, the existence of quasi-exact
solutions of the AO is also shortly illustrated.) Section \ref{Taylor}
presents a short review of the Taylor-series-based methods utilized by the
past to solve the AO. Among those is the popular Hill determinant method
(adapted to the AO in \cite{6352}). Although it is usually not classified
that way but rather among the variational methods, it is presented here
because of its close similarity with a simplistic method used sometimes to
approximately solve nonlinear ODEs \cite{3478,Fernand}. The two other
methods applied to the\ study of the linear ODE are: the power-series method 
\cite{6449}, and the analytic-continuation method \cite{6449,6432}. Then
methods with an attempt to take explicitly into account the condition at the
infinite boundary have been used: the Pad\'{e} method \cite{6201} and the
contour integral method \cite{6394}. (The interesting method developed in 
\cite{6387} which is also based on series expansion will not be mentioned
further because its generalization to any nonlinear ODE does not seem
possible.) Those latter methods consider the nonlinear form of the ODE which
induces a non-essential singularity at infinity. In section \ref{Mapping}
the principle of the mapping method is first briefly reminded. After the
transformation of the original linear ODE into a nonlinear ODE, several
configurations of the AO are explicitly treated with the mapping method in
view to illustrate its use in different cases. For $\beta =1$ and $m=2$ or $%
\beta =0$ and $m=2,3$, highly accurate estimates of the eigenvalues are
easily obtained. For double-well configurations, the accuracy decreases as $%
\beta $ is taken more and more negative (the number of terms in the Taylor
series being fixed). The phenomenon is related to the decreasing of the
radius of convergence of the Taylor series in powers of $x$ of the
logarithmic-derivative of the wave function. However the method still works,
in practice more and more terms in the series should be considered to
improve the estimates when $\beta $ takes on larger and larger negative
values. A conclusion is presented in section \ref{Conc}. Some numerical
results are displayed in the appendix.

\section{The ODE of the anharmonic oscillator\label{ODEAO}}

\subsection{The linear ODE}

The usual eigenvalue problem associated with the AO is to find the infinite
set of values $E_{n}$ which correspond to the solutions $\psi _{n}\left(
x\right) $ of the following linear ODE: 
\begin{equation}
\psi ^{\prime \prime }\left( x\right) +\left( E-V\left( x\right) \right)
\psi \left( x\right) =0  \label{eq:schro00}
\end{equation}%
with the condition that the $\psi _{n}\left( x\right) $ vanish at infinity
so as to be squared integrable. In (\ref{eq:schro00}) a prime denotes a
derivative with respect to $x$ and the potential $V\left( x\right) $
considered in the present study is:%
\begin{equation}
V\left( x\right) =\beta x^{2}+\lambda x^{2m}  \label{eq:V}
\end{equation}%
with $m=2,3,...$, and $\beta $ a given real number.

With the potential (\ref{eq:V}), the general solution of (\ref{eq:schro00})
is analytic on the whole complex $x$-plane with an irregular singular point
located at infinity.

In the following $\lambda $ can be set equal to unity without loss of
generality since the energy level of order $n$, $E_{n}^{m}\left( \beta
,\lambda \right) $, satisfies the property:%
\begin{equation*}
E_{n}^{m}\left( \beta ,\lambda \right) =\lambda ^{1/(m+1)}E_{n}^{m}\left(
\beta \lambda ^{-2/(m+1)},1\right)
\end{equation*}

The two-boundary conditions on the corresponding wave functions $\psi
_{n}\left( x\right) $ are:

\begin{itemize}
\item[1a] $\psi _{n}\left( x\right) $ are either odd or even functions of
the independent variable $x$, i. e. either $\psi _{n}\left( 0\right) =0$ or $%
\psi _{n}^{\prime }\left( 0\right) =0$.

\item[1b] $\psi _{n}\left( x\right) \rightarrow 0$ as $x\rightarrow \infty $.
\end{itemize}

For given $\beta $ and $m$, the general solution involves three arbitrary
constants: two integration constants, which are fixed by the parity
condition (1a) and a global free normalization (as it must for an
homogeneous equation), and the yet unknown "energy" parameter $E$. The
infinite set of values $E_{n}$ [the spectrum of (\ref{eq:schro00})] is then
determined by imposing the condition (1b). Actually, without this condition
and for large $x,$ the general solution of (\ref{eq:schro00}), satisfying
the condition (1a), has the following form:%
\begin{equation}
\psi \left( x\right) \overset{\scriptstyle x\rightarrow +\infty }{\simeq }%
e^{-\frac{x^{m+1}}{m+1}}+Be^{+\frac{x^{m+1}}{m+1}}  \label{eq:asygene}
\end{equation}%
where $B$ is an arbitrary constant (depending on $E$).

As a matter of fact, determining the spectrum of (\ref{eq:schro00}) amounts
to finding the (infinite) discret set of values $E_{n}$ ($n=0,1,\cdots
,\infty $) for which $B$ vanishes.

Equation (\ref{eq:asygene}) shows that when $B<0$ the general solution
vanishes at a positive large real value of $x$. Hence, values of $E\neq
E_{n} $ exist for which $\psi \left( x\right) $ vanishes on the positive
part of the real $x$-axis.

%*************************************************************************
% Figure 1 - Zeroes
\begin{figure}[tbp]
\begin{center} %viewport=200 200 1000 1200,width=10cm
\includegraphics*[viewport=-20 60 550 550,width=10cm,clip]{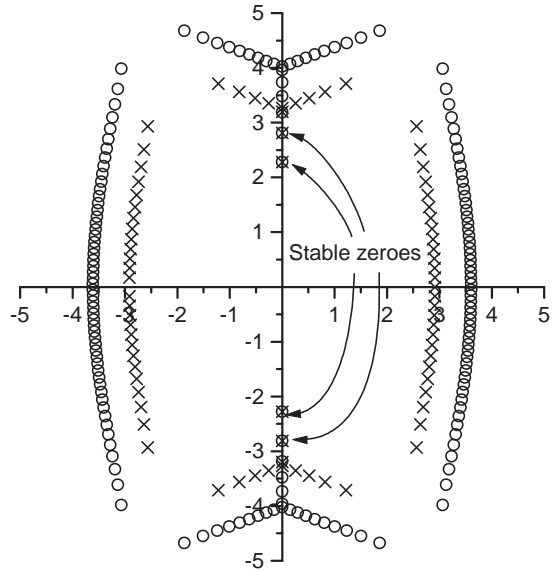}
\end{center}
%\vskip-3cm
\caption{ Zeroes in the complex $x$-plane of an approximation of the wave function $%
\protect\psi _{0}\left( x\right) $ of the fundamental state when $\protect%
\beta =1,$ $m=2$ with $E=1.39235164153029$ (close to the true value $E_{0}$,
see the appendix) and for two values of the order $M$ of the truncated
Taylor series [eq. (\protect\ref{eq:seriesPsi}) with $x=\protect\sqrt{z}$].
The approximation of $\protect\psi \left( x\right) $ is obtained by simply
summing that series. Crosses correspond to $M=40$ and open circles to $M=80$%
. One sees that the \ future (stable) zeroes of the "true" wave function $%
\protect\psi _{0}\left( x\right) $, are all probably isolated and located on
the imaginary axis. The other zeroes are potentially not properties of the
\textquotedblleft true\textquotedblright\ fundamental state, they probably
go to infinity as the order $M$ increases and as the parameter $E$
approaches the true value $E_{0}$. One observes among them the possible
existence of a positive real zero (which thus goes to $\infty $ in the same
limit).}
\label{fig1}
\end{figure}
%*************************************************************************

Figure \ref{fig1} illustrates the above considerations.

\subsection{The nonlinear ODE}

In order to remove the essential singularity at infinity displayed in (\ref%
{eq:asygene}), it is worth considering the following change of function:%
\begin{equation}
\psi \left( x\right) =x^{\epsilon }\mathrm{e}^{W\left( x\right) }
\label{eq:MansRicatti}
\end{equation}%
in which $\epsilon =0$ or $\epsilon =1$ according to the parity of the
solution of (\ref{eq:schro00}) looked for. Then the original ODE (\ref%
{eq:schro00}) becomes equivalent to:%
\begin{equation}
E-V\left( x\right) +W^{\prime 2}+W^{\prime \prime }+\frac{2\epsilon }{x}%
W^{\prime }=0  \label{eq:mansfield}
\end{equation}%
with the boundary conditions:%
\begin{equation}
W\left( 0\right) =0,\qquad W\left( x\right) \overset{x\rightarrow +\infty }{%
\simeq }-\frac{x^{m+1}}{m+1}  \label{eq:CondW}
\end{equation}

If instead of $W\left( x\right) $ one considers the derivative:%
\begin{equation}
h\left( x\right) =W^{\prime }\left( x\right)  \label{eq:h}
\end{equation}%
then (\ref{eq:mansfield}) takes the form of a Ricatti ODE (a nonlinear first
order ODE). More important is the fact that: 
\begin{equation}
h\left( x\right) =\psi ^{\prime }\left( x\right) /\psi \left( x\right)
-\epsilon /x  \label{eq:RicTransf}
\end{equation}%
so that any zero of $\psi \left( x\right) $ in the complex $x$-plane becomes
a pole for $h\left( x\right) $. For example, the nodes of the excited states 
$\psi _{n}$ ($n\geq 2$) of the AO become poles for $h\left( x\right) $
located on the positive real $x$-axis$.$

The zeroes displayed in figure (\ref{fig1}) become also poles for $h\left(
x\right) $. One sees that in the circumstances the true $h_{0}\left(
x\right) $ would be a meromorphic function of $x$ \ since it would have only
isolated poles located on the imaginary axis of the complex $x$-plane. Among
the other poles, some are present due to the approximation used (a truncated
series). But other poles are associated with the existence of a moving
singularity [characteristic of the general solution of (\ref{eq:mansfield})]
located on the positive real part of the $x$-axis and which, according to (%
\ref{eq:mansfield}, \ref{eq:h}) is of the form:

\begin{equation*}
h_{\mathrm{Sing}}\left( x\right) =\frac{1}{x_{0}-x}
\end{equation*}%
where $x_{0}$ is an arbitrary constant (the value of which depends on the
initial conditions and of $E$).

Determining the true value $E_{0}$ of the energy of the fundamental state
amounts to pushing this moving singularity toward infinity. Of course, the
location of this singularity is directly related to the arbitrary constant $%
B $ of (\ref{eq:asygene}).

Via a logarithmic-derivative-like transform (\ref{eq:RicTransf}), one is
face to the problem of solving a relatively simple nonlinear ODE the
analytic properties of which are well controlled by those of $\psi \left(
x\right) $ solution of the linear second order ODE (\ref{eq:schro00}). This
gives the opportunity of studying how the various quasi-analytical methods
for solving nonlinear ODE (presented below) actually work. In general, the
estimates of the two first energy states $E_{0}$ and $E_{1}$ of (\ref%
{eq:schro00}) obtained after a logarithmic-derivative-like transform are
more accurate than that obtained from a direct study of (\ref{eq:schro00}).
This is, presumably, due to the fact that the essential singularity located
at infinity has been soften.

\subsection{Generalities on the Taylor-series-based methods}

The methods for solving ODE based on expressing the generic solution (i.e.,
expressed in terms of the unknown $E$) as a Taylor series rely upon the
following considerations.

Consider the ODE (\ref{eq:schro00}) as an illustration. It is convenient to
perform the change of variable $x\rightarrow z=x^{2}$ and to redefine the
wave function as:%
\begin{equation}
\psi \left( x\right) =x^{\epsilon }f\left( x^{2}\right)  \label{eq:Psi}
\end{equation}%
in which $\epsilon =0$ or $\epsilon =1$ according to the parity of the
solution of (\ref{eq:schro00}) looked for. Then the original ODE (\ref%
{eq:schro00}) becomes equivalent to:

\begin{equation}
4\,z\,f^{\prime \prime }(z)+\,2\left( 1+2\,\epsilon \right) \,f^{\prime
}(z)+\,\left( E-\tilde{V}(z)\right) f(z)=0  \label{eq:Schro02}
\end{equation}%
with:%
\begin{equation*}
\tilde{V}(z)=\beta z+z^{m}
\end{equation*}%
and the boundary conditions:%
\begin{eqnarray}
f\left( 0\right) &=&1  \label{eq:0} \\
f\left( \infty \right) &=&0  \label{eq:infty}
\end{eqnarray}

Express $f\left( z\right) $ as a truncated Taylor series about the origin $%
z=0$:%
\begin{equation}
f_{M}\left( z\right) =\sum_{i=0}^{M}a_{i}\,z^{i}  \label{eq:seriesPsi}
\end{equation}%
in which $a_{0}=1$ in agreement with (\ref{eq:0}).

For $f_{\infty }\left( z\right) $ to be a generic solution of (\ref%
{eq:Schro02}), the coefficients $a_{i}$ must satisfy the following
recurrence relation [obtained by identifying to zero the coefficient of the
order $z^{k}$ in the expansion of equation (\ref{eq:Schro02})]:

\begin{equation}
\,2\left( k+1\right) \left[ 2\,k+1+2\,\epsilon \right] a_{k+1}+Ea_{k}-\,%
\beta a_{k-1}-a_{k-m}=0  \label{eq:RecurTaylor0}
\end{equation}%
with the writing convention that $a_{i}=0$ if $i<0.$

Using the recurrence relation (\ref{eq:RecurTaylor0}) one easily determines
iteratively the coefficients $a_{i}\left( E\right) $ so that $f_{\infty
}\left( z\right) $ is a solution of (\ref{eq:Schro02}) for arbitrary $E$.
The coefficicents $a_{i}\left( E\right) $ so determined are polynomials in $%
E $.

Similar considerations stand for the nonlinear ODE (\ref{eq:mansfield}) and
the expansion:%
\begin{eqnarray}
W\left( x\right) &=&g\left( x^{2}\right)  \label{eq:gTaylor0} \\
g_{M}\left( z\right) &=&\sum_{i=0}^{M}c_{i}\,z^{i}  \label{eq:gTaylor}
\end{eqnarray}%
in which $c_{0}=0$ in agreement with (\ref{eq:CondW}). The corresponding
recurrence relation is:%
\begin{equation}
2k\left( 2k-1+2\epsilon \right) c_{k}+4\sum_{i=0}^{k-2}\left( i+1\right)
\left( k-i-1\right) c_{i+1}c_{k-i-1}=\beta \,\delta _{k,2}-E\,\delta
_{k,1}+\delta _{k,m+1}  \label{eq:RecurMans}
\end{equation}%
in which $\delta _{i,j}$ is the Kronecker delta symbol. From this relation
one can determine iteratively the coefficients $c_{k}$ as polynomial
functions of $E$.

The expansion of the function $h\left( x\right) $ corresponding to the
logarithmic-derivative-like transform (\ref{eq:RicTransf}) is obtained from (%
\ref{eq:gTaylor}, \ref{eq:RecurMans}) using:%
\begin{equation*}
h_{M}\left( x\right) =2x\,g_{M}^{\prime }\left( z\right)
\end{equation*}%
with $z=x^{2}$.

To determine the set of values $E_{n}$ which potentially correspond to the
wave functions $\psi _{n}\left( x\right) $ (satisfying the boundary
condition 1b), one has to find one auxiliary condition on the coefficients $%
a_{i}\left( E\right) $ or $c_{i}\left( E\right) $. The various methods
described in sections \ref{Taylor} and \ref{Mapping} differ by the choice of
the auxiliary condition.

\subsection{Quasi-exact configurations\label{Quasi}}

Let stray from the point a little to stress on the existence of quasi-exact
configurations. Actually, for certain values of the parameters $\left( \beta
,m\right) $ a part of the spectrum of the AO may be exactly determined \cite%
{6457}. Those exact solutions express as polynomials in the independent
variable $x$ times an exponential. The existence of such solutions has been
discovered using Taylor-series-based method (Singh et al \cite{6392}, see
also \cite{6460}). It may be interesting to show the mechanism on the
particular case $m=3$.

If one factorizes the large $x$ behavior of the decreasing part of (\ref%
{eq:asygene}) before expanding in powers of $x$, then (\ref{eq:Psi}) is
replaced by:%
\begin{equation}
\psi \left( x\right) =x^{\epsilon }e^{-\frac{x^{4}}{4}}\tilde{f}\left(
x^{2}\right)
\end{equation}%
and (\ref{eq:RecurTaylor0}) becomes a three-terms recurrence relation:

\begin{equation}
2\,\left( 1+k\right) \,\left( 2\,k+1+2\,\epsilon \right) \tilde{a}_{k+1}+E\,%
\tilde{a}_{k}-\left( 4\,k+2\,\epsilon -1+\beta \right) \tilde{a}_{k-1}=0
\label{eq:RecurQ}
\end{equation}

Then, if two successive coefficients vanish, the series terminates to give a
polynomial. The corresponding solutions are exactly determined. One easily
obtains, from (\ref{eq:RecurQ}), the following exact configurations for $%
\epsilon =0$: 
\begin{eqnarray*}
&&\beta =-3,\quad E=0,\quad \tilde{f}\left( z\right) =1 \\
&&\beta =-7,\quad E=\pm 2\sqrt{2},\quad \tilde{f}_{\pm }\left( z\right)
=1\mp \sqrt{2}z \\
&&\beta =-11,\quad \left\{ 
\begin{array}{l}
E=\pm 8,\quad \tilde{f}_{\pm }\left( z\right) =1\mp 4z+2z^{2}, \\ 
E=0,\quad \tilde{f}\left( z\right) =1-\frac{2}{3}z^{2}%
\end{array}%
\right. \\
&&\vdots
\end{eqnarray*}%
and for $\epsilon =1$: 
\begin{eqnarray*}
\beta &=&-5,\quad E=0,\quad \tilde{f}\left( z\right) =1 \\
\beta &=&-9,\quad E=\pm 2\sqrt{6},\quad \tilde{f}_{\pm }\left( z\right)
=1\mp \sqrt{\frac{2}{3}}z \\
&&\vdots
\end{eqnarray*}

There is one exact configuration for each value of $\beta $ for which the
coefficient of $a_{k-1}$ in (\ref{eq:RecurQ}) vanishes (provided $k>1)$,
namely for the values:%
\begin{equation*}
\beta _{j}=-\left( 4j+3+2\epsilon \right) ,\quad j=0,\,1,\,2,\,3,\,\cdots
\end{equation*}%
For each of these values, there are $j+1$ exactly soluble eigenvalues
whereas the remaining eigenvalues of the complete spectrum cannot be exactly
determined. In section \ref{double}, the quasi-exact configuration $\beta
=-11$ is explicitly treated with the mapping method.

\section{The Taylor-series-based methods used to solve the AO \label{Taylor}}

\subsection{The power-series method\label{Power}}

The analyticity properties of the solutions of (\ref{eq:Schro02}) suggest a
very simple method for solving the eigenvalue problem of (\ref{eq:schro00}),
namely the power-series method \cite{6449}.

Using the property that the Taylor series $f_{\infty }\left( z\right) $ has
an infinite radius of convergence, one may determine the infinite set of
values of $E$ for which $f_{\infty }\left( z_{0}\right) $ vanishes at a
given finite $z_{0}$ by simply imposing that the sum of the series (\ref%
{eq:seriesPsi}) vanishes at this point. This zero located on the positive
part of the real axis of $z$ is associated with a general solution of (\ref%
{eq:Schro02}) corresponding to the asymptotic form (\ref{eq:asygene}) with $%
B\neq 0$. Then, the limit $z_{0}\rightarrow \infty $ should correspond to $%
B\rightarrow 0.$ This provides the possibility of approaching the spectrum $%
E_{n}$ of the AO (as actually observed in several studies of "bounded"
oscillators \cite{6444}).

Dealing with finite values of $M$, $f_{M}\left( z\right) $ represents
correctly $f_{\infty }\left( z\right) $ only in a finite range of values of $%
z$ whereas for larger values the sum goes to $\pm \infty $ according to the
sign of the last term. The procedure for determining the spectrum of the AO
may then be sketched as follows.

\begin{enumerate}
\item[2a] One fixes $M$, chooses a consequent but not too large value of $%
z_{0},$ and determines the zeroes of the polynomial in $E$ associated to the
solution of the equation 
\begin{equation}
f_{M}\left( z_{0}\right) =0  \label{eq:CondPower}
\end{equation}%
(this is the effective auxiliary condition at $z_{0}$).

\item[2b] On increasing $z_{0}$ one observes a convergence of the zeroes
towards definite values up to a value $z_{0}^{(M)}$ where the convergence
terminates. At this point one gets the best values of the spectrum $%
E_{n}\left( z_{0}^{\left( M\right) }\right) $ for the given $M$.

\item[2c] One increases $M$ and repeats the process from step 2a.

\item[2d] One observes that $z_{0}^{(M)}$ is an increasing function of $M$
and that $E_{n}\left( z_{0}^{\left( M\right) }\right) $ quickly approaches
the spectrum $E_{n}$ as $M$ grows.
\end{enumerate}

This method has been applied in 1961 by Secrest et al \cite{6449} with $%
M\sim 1000$. They notably have determined the fundamental energy $E_{0}$
with 12 significant figures in the case $m=2$, $\beta =0$. This method has
been tried again during the present work, for $m=2$, $\beta =1$ and $M=250$
it gives the fundamental energy $E_{0}$ with 41 significant figures!
%*************************************************************************
% Figure 2 - Power series
\begin{figure}[tbp]
\begin{center}
\includegraphics*[viewport=20 40 400 600,width=10cm,clip]{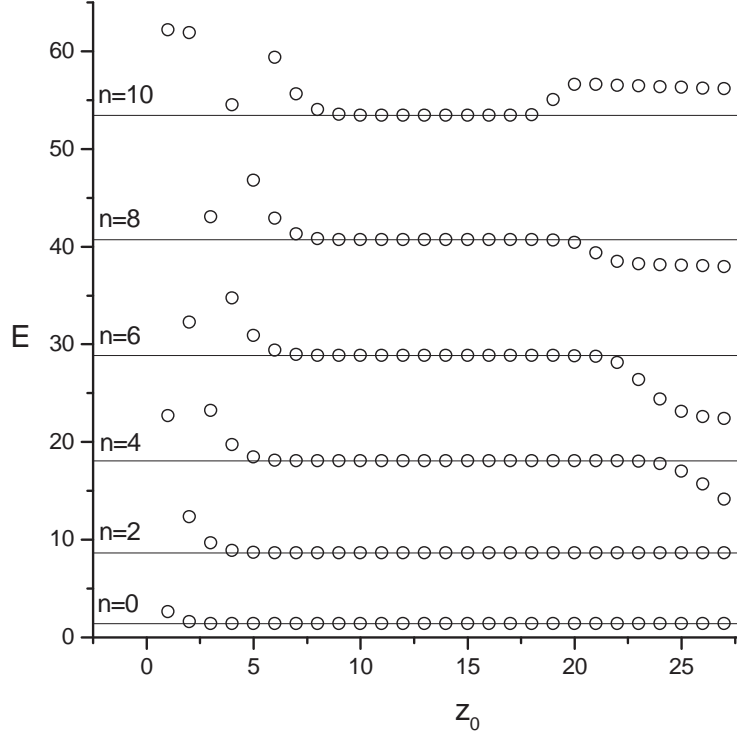}
\end{center}
\vskip-3cm
\caption{ Part of the even spectrum obtained
using the power series method for $\protect\beta =1,$ $m=2$ and with $M=100$%
. The open circles correspond the real zeroes of the polynomial condition (%
\protect\ref{eq:CondPower}) for varying $z_{0}$. The horizontal lines
indicate the \textquotedblleft true\textquotedblright\ values of the
spectrum ($E_{0}$, $E_{2}$, $E_{4}$, $\cdots $) as displayed in the
appendix. For sufficiently high values of $z_{0}$ the points deviate from
the lines because $f_{M}\left( z\right) $ no longer represents correctly the
function $f_{\infty }\left( z\right) $. One sees also that the larger the
quantum number $n$ is, the larger the minimal value of $z_{0}$ required to
identify the existence of the energy state must be.}
\label{fig2}
\end{figure}
%*************************************************************************

As illustrated by figure (\ref{fig2}), the determination of all the real
zeroes of the condition (\ref{eq:CondPower}) provides an estimation of the
complete spectrum, the number of states being only limited by the order of
the polynomial in $E$ corresponding to the value of $M$ chosen. The excited
states may be hierarchically determined that way provided $z_{0}$ is chosen
larger than the\ location of the last node. As a consequence the number of
significant figures obtained slowly decreases as the quantum number $n$
increases (for a given $M$).

It may occur (for some values of the potential parameters $\beta $ and $m)$
that too large values of $z_{0}$ or $M$ be required to get a satisfactory
accuracy. In that case Secrest et al \cite{6449} have suggested to reach the
value $z_{0}$ in more than one step using the Taylor expansion about a
non-zero value $z_{l}<z_{0}$. This suggestion is already the
analytic-continuation method introduced later on by Holubec and Stauffer 
\cite{6432}.

\subsection{The analytic-continuation method\label{Continuation}}

Let us consider the Taylor expansion of $f\left( z\right) $ about an
arbitrary point $z_{l}\neq 0$:%
\begin{equation}
f_{M,l}\left( z\right) =\sum_{i=0}^{M}b_{i}\,\left( z-z_{l}\right) ^{i}
\label{eq:serieskPsi}
\end{equation}

For $f_{\infty ,l}\left( z\right) $ to be a generic solution of (\ref%
{eq:Schro02}), the coefficients $b_{i}$ must satisfy the following
recurrence relation:%
\begin{eqnarray}
4\,z_{l}\,\left( k+2\right) \left( k+1\right) b_{k+2} &=&\,2\left(
k+1\right) \left[ 2k+1+2\,\epsilon \right] b_{k+1}  \notag \\
&&+\,\left( E-\beta z_{l}-z_{l}^{m}\right) b_{k}-\left( \beta
+mz_{l}^{m-1}\right) b_{k-1}  \notag \\
&&-\sum_{i=0}^{m-2}\frac{m!}{i!(m-i)!}z_{l}^{i}b_{k-m+i}  \label{eq:recur2}
\end{eqnarray}%
which determines them iteratively in terms of $E$ and of the two arbitrary
coefficients $b_{0}$ and $b_{1}$. Those two quantities may be calculated
using the series at the origin (\ref{eq:seriesPsi})\ so as to account for
the conditions 1a at the origin with thus:%
\begin{equation}
b_{0}=f_{\infty }\left( z_{l}\right) ,\qquad b_{1}=f_{\infty }^{\,\prime
}\left( z_{l}\right)  \label{eq:b0b1}
\end{equation}

For a given $M$, having chosen, for example, $z_{l}=z-z_{l}=h$ to be small,
one may estimate $f_{M}\left( h\right) $ and $f_{M}^{\prime }\left( h\right) 
$ that way and then $f_{M}\left( 2h\right) $ as a function of the unique
unknown parameter $E$ using (\ref{eq:serieskPsi}--\ref{eq:b0b1}). One uses
that result to estimate again $b_{0}$ and $b_{1}$ at this point in order to
reach the point $3h$ and so on until the point $z_{0}=N\,h$ is reached.

The following of the procedure is identical to the preceding one (steps
2a--2d above) but is relatively slow. Nevertheless it is very efficient
provided a well balanced choice of $M$, $N$ and $h$ is done. For example,
with $M=N=40$, $22$ significant figures have been obtained this way \cite%
{6431} in the case of the fundamental energy of the double-well potential $%
\beta =-50$ and $m=2$ for which it is almost degenerated with the first
excited state (the splitting occurs at the $21^{st}$ figure only). In the
circumstances, it is more efficient than the mapping method (see section \ref%
{double}).

Notice that, contrary to the power-series method, the analytic-continuation
method could apply even in configurations where the range of analyticity of $%
\psi $ about the origin is limited.

\subsection{The Hill determinant method}

One of the most popular methods used to solve the eigenvalue problem of the
AO is the so-called Hill determinant method \cite{6352} which is (partly)
based on a Taylor expansion of the solution in powers of the independent
variable $x$ such as (\ref{eq:Psi}, \ref{eq:seriesPsi}).

Actually it is traditionally considered not as a pure power-series-based
method but rather as a variational method because of an exponential
prefactor usually introduced in the relation (\ref{eq:Psi}) between $\psi $
and the power series (\ref{eq:seriesPsi}). In \cite{6352}, Biswas et al use
a fixed exponential factor $e^{-x^{2}/2}$ but subsequent studies considered $%
e^{-\gamma x^{2}}$ with $\gamma $ adjustable \cite{6417,6358,6348} and
even $e^{-\gamma x^{2}+\rho x^{4}}$ with $\gamma $ and $\rho $ adjustable 
\cite{6392}. Nevertheless this method may well be sketchily introduced
without considering any prefactor. That will evidence a similitude with a
simplistic method (see section \ref{simplistic}) used sometimes \cite{3478}
to solve nonlinear ODEs with the help of power series.

Let us consider the recurrence relation (\ref{eq:RecurTaylor0}) as an
infinite system of linear algebraic equations for the coefficients $a_{i}$.
For this homogeneous linear system, to have a solution, its (infinite)
determinant (name the Hill determinant, see e.g. \cite{whitt}), if it
converges, must vanish. This condition is a transcendental equation for $E$
the infinite number of solutions of which should coincide with the complete
spectrum of the AO provided the condition at infinity is satisfied.

Of course, practically, one deals with finite values of $M$ and the
effective auxiliary condition reduces to:%
\begin{equation}
D_{M}\left( E\right) =0  \label{eq:HillCondM}
\end{equation}%
where $D_{M}\left( E\right) $ is the Hill determinant truncated at order $M$.

In fact, the auxiliary condition (\ref{eq:HillCondM}) is not always
sufficient to determine the solution looked for. This is due to the
analyticity of the general solution of (\ref{eq:schro00}): the blowing part
of the solution also corresponds to a convergent series and may be selected
by the iterative procedure so defined. In practice one should verify that
the wave function selected by the Hill criterium (\ref{eq:HillCondM})
actually vanishes at infinity.

Using an exponentially decreasing prefactor one may improve the method. It
is a matter of fact that this \textquotedblleft genuine\textquotedblright\
Hill determinant method works in certain circumstances (see for example \cite%
{6348}). But it fails in certain other circumstances. Especially it is
unable to furnish the full spectrum of the quasi-exact configurations \cite%
{6478} like those presented in section \ref{Quasi}, see \cite{6463} for a
review of the reasons of failure.

\subsection{A simplistic method\label{simplistic}}

At this stage, it is interesting to realize that:%
\begin{equation*}
a_{M+1}\left( E\right) =(-1)^{M+1}\frac{D_{M}\left( E\right) }{d_{M+1,0}}%
a_{0}
\end{equation*}%
where $d_{M+1,0}$ is the $\left( M+1,0\right) $\ minor of the matrix of the
system of linear equations (\ref{eq:RecurTaylor0}). Hence the condition (\ref%
{eq:HillCondM}) is equivalent to the condition that the next coefficient $%
a_{M+1}(E)$ vanishes (e.g. see \cite{6392,6441}). This is the condition
imposed in certain studies of nonlinear ODEs, thus assuming implicitly that
the Taylor series converges. Unfortunately, in general, this is not true 
\cite{3553,Fernand}, due to the presence of singularities in the complex $x$%
-plane \cite{3358} in the case of a nonlinear ODE.

For example, the solutions of the nonlinear ODE (\ref{eq:mansfield}) display
singularities which limit the convergence of the Taylor series (\ref%
{eq:gTaylor}). In the case corresponding to the logarithmic derivative of
the wave function the zeroes of which are displayed in figure \ref{fig1} ($%
\beta =1,$ $m=2$), one may estimate the radius of convergence of its Taylor
series from the knowledge of the location of the singularity the closest to
the origin (which is located on the imaginary axis of the complex $x$-plane,
see figure \ref{fig1}). In terms of $z=x^{2}$ one obtains a radius of
convergece $\left\vert z\right\vert <R_{0}\simeq 5.192695$. Now by imposing
the auxiliary condition:%
\begin{equation}
c_{M}\left( E\right) =0  \label{eq:simplistic}
\end{equation}%
one assumes that the series converges. But this is only true if $\left\vert
z\right\vert <R_{0}$.

The condition (\ref{eq:simplistic}) is not sufficient to impose the required
condition at infinity, then it does not yield a convergent procedure for
subsequent values of $M$. Nevertheless it, sometimes, gives approximate
estimates of the true spectrum. This is because the degree of accuracy on
the determination of the spectrum is associated with the location of the
moving singularity. If a large accuracy on $E_{0}$ is required to push the
moving singularity beyond the range of convergence of the series then one
may expect that (\ref{eq:simplistic}) will give some (limited) information
on $E_{0}$. In the contrary case, the procedure simply does not work. This
phenomenon may explain why the simplistic method "works" in some case and
does not in another case despite similar radius of convergence for the
respective Taylor series \cite{6137}.

This simplistic method, when it works, is extremely easy to implement and
may serve to get a primary estimate of the spectrum before using more
sophisticated methods such as the following ones which are attempts to
construct an auxiliary condition that accounts explicitly for the asymptotic
behavior of the solution looked for. To this end, the singularity located at
infinity must be soften, hence these methods apply to the nonlinear ODE (\ref%
{eq:mansfield}).

\subsection{The Pad\'{e} method}

The Pad\'{e} method (originally proposed in \cite{6201}) relies upon an
attempt to represent the solution of (\ref{eq:mansfield}) looked for by
successive rational functions of the form:

\begin{equation*}
P_{N_{1},N_{2}}\left( z\right) =\frac{\sum_{i=0}^{N_{1}}p_{i}z^{i}}{%
\sum_{i=0}^{N_{2}}q_{i}z^{i}}
\end{equation*}%
which involves $N_{1}+N_{2}+1$ coefficients and $z=x^{2}$. The processus of
the method may be described as follows.

As is known, the coefficients $p_{i}$ and $q_{i}$ may be determined from the
Taylor series (\ref{eq:gTaylor}) at order $M=N_{1}+N_{2}$ according to the
usual rules of construction of a Pad\'{e} approximant. This standard
procedure determines the coefficients $p_{i}$ and $q_{i}$ as functions of
the unknown parameter $E$. The auxiliary condition is then obtained by
imposing that the Pad\'{e} approximant so constructed at order $%
M=N_{1}+N_{2} $ still reproduces the truncated function at next order $M+1$.
Namely, setting $N_{1}=s+\omega $ and $N_{2}=s$, one gets the following
linear system of equations for the coefficients $p_{i}$ and $q_{i}$ (using
the convention that $c_{k}=0$ for $k<0$): 
\begin{eqnarray*}
p_{i} &=&\sum_{j=0}^{s}c_{i-j}q_{j}\quad \mathrm{for\quad }i=1,\cdots
,s+\omega \\
0 &=&\sum_{j=0}^{s}c_{i-j}q_{j}\quad \mathrm{for\quad }i=s+\omega +1,\cdots
,2s+\omega +1
\end{eqnarray*}

The second line is an homogeneous system of linear algebraic equations for
the $s+1$ coefficients $q_{j}$. Then the condition to have a solution is
that the matrix:%
\begin{equation*}
\hat{T}_{i,j}=c_{s+\omega +1+i-j}\qquad \left( i=0,\cdots ,s;j=0,\cdots
,s\right)
\end{equation*}%
must have a vanishing determinant. In the circumstances, $\hat{T}_{i,j}$ is
a Toeplitz matrix which may be rewritten under the form of a Hankel matrix $%
\hat{H}_{i,k}$ by a redefinition of the indice $k=s-j$ with thus:%
\begin{equation*}
\hat{H}_{i,k}=c_{\omega +1+i+k}\qquad \left( i=0,\cdots ,s;k=0,\cdots
,s\right)
\end{equation*}

Finally, the condition%
\begin{equation}
\det (\hat{H})\equiv \det \left( \hat{T}\right) =0  \label{eq:detH}
\end{equation}%
is the auxiliary condition looked for to determine the spectrum provided
that one chooses $\omega $ in agreement with the boundary condition at
infinity. In general it is sufficient to choose one of the three values $%
\omega =1,0,-1$ according to whether the function to be determined goes to $%
\pm \infty $, a constant or $0$ when $z\rightarrow \infty $. Eventually,
considering two successive values of $\omega $ gives upper and lower bounds
on the eigenvalues \cite{6191,6201,6389}.

The advantage of the Pad\'{e} method is that it is well adapted to
reproduce, from its Taylor expansion, the analytic structure of a
meromorphic function especially if it has poles. Now, as explain above, the
logarithmic-derivative transform changes the zeroes of $\psi \left( x\right) 
$ into poles for $h\left( x\right) $.\ Hence the Pad\'{e} method is better
adapted to sum the Taylor series of $h\left( x\right) $ rather than that of $%
\psi \left( x\right) $. Moreover, it is also, a priori, able to determine
the energies of the excited states despite (or rather because of) the poles
located on the positive real $x$-axis (the nodes of $\psi _{n}\left(
x\right) $ for $n\geq 2$).

The Pad\'{e} method (named the Ricatti-Pad\'{e} method in \cite{6335} and
later on the Hankel-Pad\'{e} method in \cite{6194}) has been first
introduced in \cite{6201} in conjonction with a logarithmic-derivative
transform like (\ref{eq:RicTransf}) to calculate, notably, the even and odd
fundamental energies of the AO with $\beta =1$ and $m=2$ and for various
values of $\lambda $. Typically the accuracy obtained was about 8
significant figures. Some time later, in \cite{6191}, the excited state
energies have been estimated for the pure quartic and sextic AO (i.e. $\beta
=0$, $m=2$ and $3$), eleven significant figures were obtained on the
estimate of $E_{0}$ for the quartic AO. This method has then been utilized
several times (see e.g. \cite{6389} for a list of references). In \cite{6389}
the first two eigenvalues of both the quartic AO and the double well down to 
$\beta =-15$ are estimated with 20--18 significant figures. In addition,
Amore and Fern\'andez \cite{6190} have shown that the Pad\'{e} method may also
be applied to solve the two-point boundary value problem associated to
several nonlinear ODEs.

In general the Pad\'{e} method is easy to use [because of the simple writing
of the auxiliary condition (\ref{eq:detH})]. It also appears to be robust in
several occasions. However its effectiveness is limited because the
(repeated) calculations of determinants of large matrices are extremely time
consuming. Sometimes, the Pad\'{e} approximants introduce "spurious" poles
or zeroes that can perturbate a clear determination of the spectrum $E_{n}$.
Though it is not as refined as the following methods, the Pad\'{e} method
may be extremely useful (see section \ref{double}).

\subsection{The contour-integral method\label{contour}}

\subsubsection{The fundamental state}

Leonard and Mansfield \cite{6394,6398} have proposed the recourse to a
contour integral in the complex $x$-plane to perform an analytic
continuation of the Taylor series of $W\left( x\right) $ [satisfying (\ref%
{eq:mansfield})] towards the large $x$ values so that the asymptotic
behavior (\ref{eq:CondW}) can be effectively imposed. The method may be
described as follows.

Starting with the Taylor series (\ref{eq:gTaylor}) the coefficients of which
satisfy the recurrence relation (\ref{eq:RecurMans}), one rewrites this
series in terms of large $s=1/x$:%
\begin{equation}
\tilde{g}_{M}\left( s\right) =\sum_{i=0}^{M}\frac{c_{i}{}}{s^{2i}}\,
\label{eq:gtildeM}
\end{equation}%
\noindent

According to (\ref{eq:CondW}), one is interested in finding the values of $E$
for which this series, for $M\rightarrow \infty $, has a pole of order $m+1$
at the origin $s=0$:%
\begin{equation}
\tilde{g}_{\infty }\left( s\right) \overset{s\rightarrow 0}{\simeq }-\frac{1%
}{\left( m+1\right) \,s^{m+1}}  \label{eq:poleOri}
\end{equation}

To this end, one considers the following integral over a large circle
contour $C$ around the origin:%
\begin{equation*}
F_{\infty }\left( \sigma \right) =\frac{1}{2i\pi \sigma ^{m+1}}\int_{C}\frac{%
e^{\sigma s}}{s}\,\tilde{g}_{\infty }\left( s\right) \,\mathrm{d}s
\end{equation*}%
then, according to the Cauchy formula the contribution of the pole (\ref%
{eq:poleOri}) to this function is $-\frac{1}{(m+1)!\left( m+1\right) }\,$.
Assuming that all the other singularities of $\tilde{g}_{\infty }\left(
s\right) $ are located on the left of the imaginary axis in the complex $s$%
-plane then their contributions to $F_{\infty }\left( \sigma \right) $ will
be made negligible as $\sigma \rightarrow \infty $, so that:%
\begin{equation}
\lim_{\sigma \rightarrow \infty }F_{\infty }\left( \sigma \right) =-\frac{1}{%
(m+1)!\left( m+1\right) }\,  \label{eq:limite}
\end{equation}

The truncated series (\ref{eq:gtildeM}) is then used to estimate $F_{\infty
}\left( \sigma \right) $, leading to:%
\begin{eqnarray}
F_{M}\left( \sigma \right) &=&\frac{1}{2i\pi \sigma ^{m+1}}%
\sum_{i=0}^{M}c_{i}\,\int_{C}\frac{e^{\sigma s}}{s}\,\,s^{-2i}\,\mathrm{d}s 
\notag \\
&=&\sum_{i=0}^{M}\frac{c_{i}}{\Gamma \left( 2i+1\right) }\sigma ^{2i-m-1}
\label{eq:Factorielle}
\end{eqnarray}

Owing to the Euler Gamma function in the denominator, this series converges
(the original series had a finite radius of convergence) and may be summed
term by term to estimate $F_{\infty }\left( \sigma \right) $ when $\sigma $
becomes large. Then a procedure similar to that described in points 2a to 2d
in section \ref{Power} may be applied with $z_{0}$ replaced by $\sigma _{0}$
at which point the condition (\ref{eq:limite}) is tentatively imposed.

If the region of analyticity of $\tilde{g}_{\infty }\left( s\right) $ does
not correspond to the assumption that all the singularities are located on
the left of the imaginary axis of $s$, the convergence may be spoiled by
irreductible oscillations. Notice that this is the case for $\beta =1$, $m=2$
since, according to figure \ref{fig1}, the zeroes of $\psi _{0}\left(
x\right) $ are located right on the imaginary axis. Thus $W_{0}\left(
x\right) =\ln \left( \psi _{0}\left( x\right) \right) $ has singularities on
this axis. To circumvent such difficulties, Leonard and Mansfield \cite%
{6394,6398}\ propose to modify (\ref{eq:Factorielle}) by introducing a
parameter $\alpha _{I}$, so that:%
\begin{equation*}
F_{M,\alpha _{I}}\left( \sigma \right) =\sum_{i=0}^{M}\frac{c_{i}}{\Gamma
\left( 2i\alpha _{I}+1\right) }\sigma ^{2i-m-1}
\end{equation*}%
what corresponds to having performed some rotation of the complex $s$-plane
(on the left hand side if $\alpha _{I}<1$).

\subsubsection{The excited states}

Because the excited states $\psi _{n}\left( x\right) $ for $n\geq 2$ have
nodes on the positive real part of the $x$-axis, the corresponding $%
W_{n}\left( x\right) $ have singularities on the positive real part of the $%
s $-axis which cannot be moved by $\alpha _{I}$. Hence the method does not
apply directly to the determination of the excited states. In order to have
access to them, Leonard and Mansfield \cite{6394} propose to use $%
W_{0}\left( x\right) $ (determined by the procedure described just above) as
a basis to write [here accounting for the odd ($\epsilon =1$) and even ($%
\epsilon =0$) possibilities]:%
\begin{equation}
\psi \left( x\right) =x^{\epsilon }e^{W_{\epsilon }\left( x\right) }P\left(
x\right)  \label{eq:Factor}
\end{equation}

The ODE satisfied by $P\left( x\right) $ is then:%
\begin{equation*}
P^{\prime \prime }+\frac{2\epsilon }{x}P^{\prime }+2W_{\epsilon }^{\prime
}P^{\prime }+\left( E-E_{\epsilon }\right) P=0
\end{equation*}

This differential equation has two types of large $x$ solution:%
\begin{eqnarray}
&&P\left( x\right) \overset{x\rightarrow \infty }{\simeq }\exp \left[ -\frac{%
E-E_{\epsilon }}{2\left( m-1\right) x^{m-1}}\right]  \label{eq:behav1} \\
&&P\left( x\right) \overset{x\rightarrow \infty }{\simeq }\exp \left[ 2\frac{%
x^{m+1}}{\left( m+1\right) }\right]  \label{eq:behav2}
\end{eqnarray}

Only (\ref{eq:behav1}) is compatible with the boundary condition at infinity
for $\psi _{n}\left( x\right) $. Actually, because in (\ref{eq:Factor}) $%
W_{\epsilon }\left( x\right) $ already factorizes the exponential decreasing
of $\psi _{n}\left( x\right) $, $P_{n}\left( x\right) $ must approach a
constant, then (\ref{eq:behav1}) satisfies this condition since $m>1$.
Taking into account (\ref{eq:CondW}), one sees that the second behavior (\ref%
{eq:behav2}) reconstructs the blowing part\ proportional to $B$ in (\ref%
{eq:asygene}) as it must.

With the behavior (\ref{eq:behav1}), $P_{n}\left( 1/s\right) $ has no
singularity at $s=0$ and the contour integral procedure may again be applied
to determine the values of $E_{n}-E_{\epsilon }$ that make this integral
vanish when $\sigma \rightarrow \infty $.

Notice that the asymptotic condition (zero for the excited states) being
independent of the energy state, an individual adjustment may correspond to
any value $E_{n}$ so that a global (even partial) obtention of the spectrum
[as that illustrated by figure (\ref{fig2})] is not possible. In particular,
a plot of the wave function is necessary to identify the excited state so
determined \cite{6394}. One could, however, use a procedure similar to that
of section \ref{Power} for the power-series method to determine globally the
spectrum before looking for higher accuracy on the determination of each
individual energy state.

\subsubsection{Rescaling}

In order to improve the efficiency of their method applied to the AO,
Leonard and Mansfield \cite{6394,6398} utilize a rescaling which allows them
to assign to the third coefficent $c_{3}$ the role of the adjustement
parameter instead of $E$ whereas $c_{2}$ is fixed instead of $\beta $. With
this trick and 300 terms in the series ($M=300$), they are able (for $c_{2}$
fixed to zero, hence $\beta =0$ and $m=2$) to estimate $E_{0}$ \ with an
accuracy of $65$ significant figures. The remaining of the spectrum of the
AO up to $n\simeq 40$ is determined with an accuracy of $48$ significant
figures. That is a remarkable achievement.

It is to be noted that the rescaling trick is not very convenient when $%
\beta \neq 0$ since then a supplementary adjustment of $c_{2}$ is required 
\cite{6394}.

As indicated below, the mapping method appears to be more efficient than the
contour-integral method, since, using of the same rescaling trick, it yields
a much greater accuracy whereas $M=250$ only for the configuration $\beta =0$
and $m=2$ (see the appendix).

The major drawback of the contour integral method is the iterative
uncomfortable adjustment procedure necessary to approach the infinite
boundary (similar to that of the power-series method described in section %
\ref{Power}).

\section{The mapping method\label{Mapping}}

\subsection{Introduction}

%*************************************************************************
% Figure 3 - Contour
\begin{figure}[tbp]
\begin{center} %viewport=200 200 1000 1200,width=10cm
\includegraphics*[viewport=-20 60 550 550,width=10cm,clip]{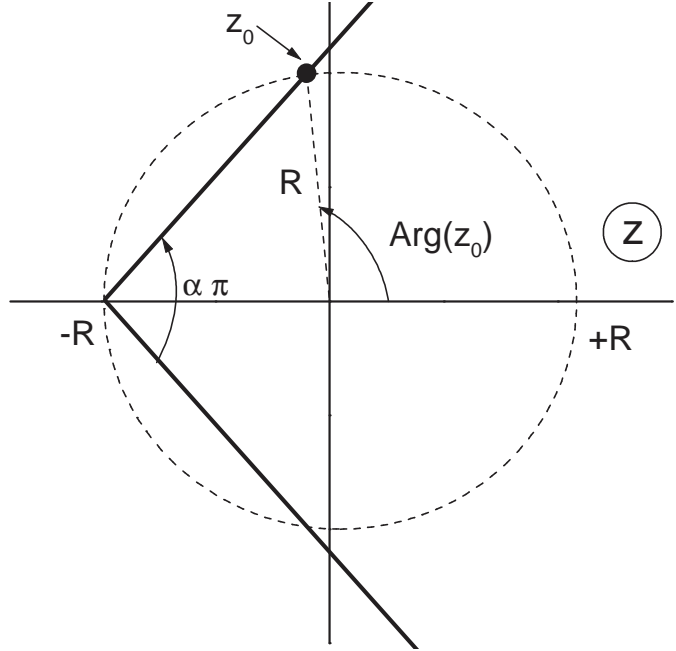}
\end{center}
%\vskip-3cm
\caption{When the interior of an angular
sector involving the positive real axis of $z$ defines a region of
analyticity for the function $f\left( z\right) $ (with $z=x^{2}$), it may be
mapped onto the interior of a circle of unit radius for the complex variable 
$w$ defined by eq (\protect\ref{eq:Map}). The singularity $z_{0}$ of $%
f\left( z\right) $ the closest to the origin determines the analyticity
domain provided the other singularities lay outside the angular sector
containing the origin.}
\label{fig3}
\end{figure}
%*************************************************************************

In solving the AO problem, it is important to take correctly the condition
at infinity (an essential singularity) into account. It is thus useful to
first reduce this singularity. To this end a logarithmic-derivative of the
wave function is usually performed. As illustrated with the Pad\'{e} and the
contour-integral methods, the condition at infinity may then be more
efficiently imposed. With the contour integral method an analytic
continuation of the original Taylor series towards the infinite boundary is
used. But no attempt is made to compactify the integration domain of the
original ODE, so that the practical execution is finally similar to that of
the analytic continuation method (the power-series method even) with a
clumsy step-by-step approach to a point located at infinity.

The mapping method introduced in \cite{6319} leans also on an analytic
continuation after a logarithmic-derivative-like transformation but in
addition, the infinite boundary is brought close to the origin using the
following conformal transformation:

\begin{equation}
z\rightarrow w=\frac{\left( 1+z/R\right) ^{1/\alpha }-1}{\left( 1+z/R\right)
^{1/\alpha }+1}\,,  \label{eq:Map}
\end{equation}%
in which $R$ and $\alpha $ characterize the position of the vertex and the
angle of an angular sector of the complex plane of $z$ as shown in figure %
\ref{fig3}.

The conformal transformation (\ref{eq:Map}) maps the interior of the angular
sector of the $z$-plane into the interior of the unit circle centered at the
origin of the $w$-plane so that $z=\infty $ corresponds to $w=1$ (whereas $%
z=0$ corresponds to $w=0$).

If the interior of the angular sector is a region of analyticty of the
original function $g\left( z\right) $ then the series (\ref{eq:gTaylor})
converges there. Under the analytic continuation this convergence is
conveyed to the whole unit disc for the series: 
\begin{equation}
\tilde{g}_{M}\left( w\right) =\sum_{i=0}^{M}u_{i}\left( E\right) \,w^{i}
\label{eq:SerieAfterMap}
\end{equation}
obtained by expanding, within the original series (\ref{eq:gTaylor}), the
relation inverse of (\ref{eq:Map}):%
\begin{equation}
z=R\left[ \left( \frac{1+w}{1-w}\right) ^{\alpha }-1\right] \,.
\label{eq:MapInverse}
\end{equation}

\subsection{Example of the AO with $\protect\beta =1$ and $m=2$}

%*************************************************************************
% Figure 5 - Convergence
\begin{figure}[tbp]
\begin{center} %viewport=200 200 1000 1200,width=10cm
\includegraphics*[viewport=-20 60 550 550,width=15cm,clip]{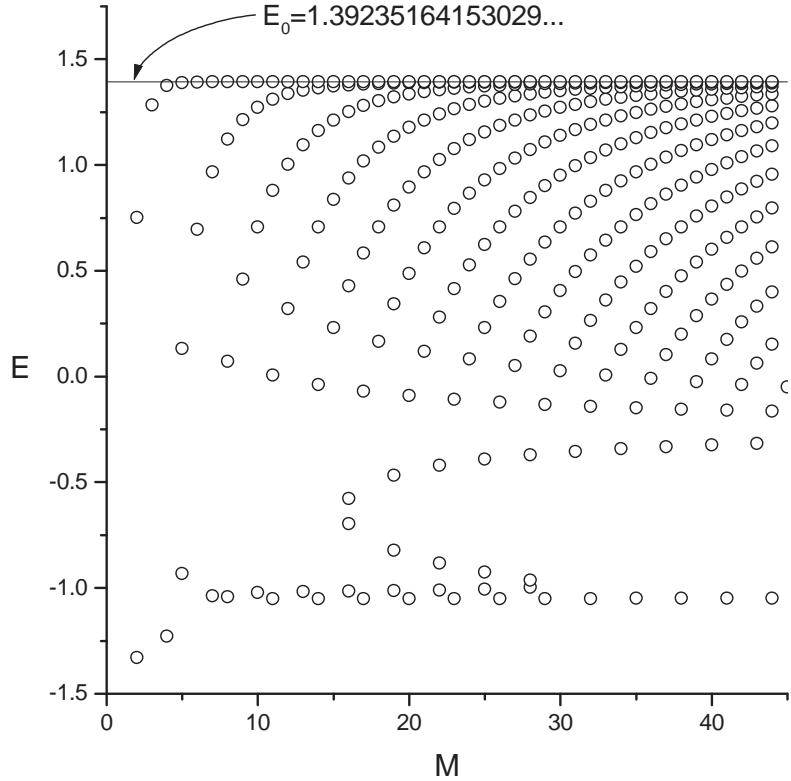}
\end{center}
\vskip-2.5cm
\caption{Evolution with the order $M$ of
all the real zeroes corresponding to the auxiliary condition (\protect\ref%
{eq:g2w1}) for the AO with $\protect\beta =1,$ $m=2$. The mapping transform (%
\protect\ref{eq:Map})\ is used with $R=5.192695$ and $\protect\alpha =2$
corresponding to the expected analytic structure of the fundametal state $%
\protect\psi _{0}$ (see figure \protect\ref{fig1} and text). The candidate
to the fundamental energy $E_{0}$ is well evidenced.}
\label{fig5}
\end{figure}
%*************************************************************************

%*************************************************************************
% Figure 4 - Convergence
\begin{figure}[tbp]
\begin{center} %viewport=200 200 1000 1200,width=10cm
\includegraphics*[viewport=-20 60 550 550,width=15cm,clip]{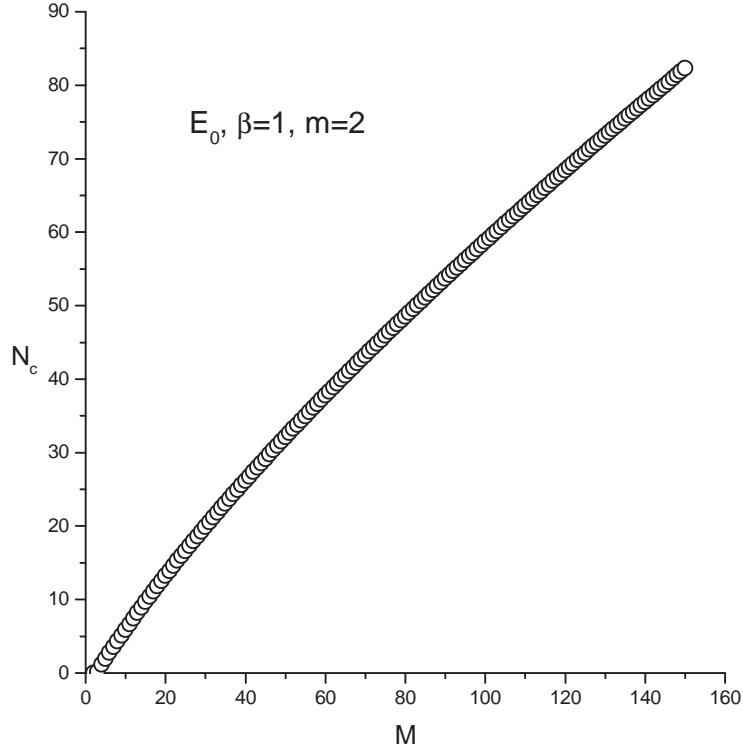}
\end{center}
\vskip-2cm
\caption{Variation with the
order $M$ of the approximate number of stabilized figures ($N_{\mathrm{c}%
}=-\log \left\vert 1-E_{M}/E_{M-1}\right\vert $ where $E_{M-1}$ and $E_{M}$
are two successive estimates of $E_{0}$) in the determination of the
fundamental energy $E_{0}$ of the AO with $\protect\beta =1,$ $m=2$. The
method used is the conformal mapping of (\protect\ref{eq:Map}) with $%
R=5.192695$ and $\protect\alpha =2$.}
\label{fig4}
\end{figure}
%*************************************************************************

%*************************************************************************
% Figure 6 - Convergence
\begin{figure}[tbp]
\begin{center} %viewport=200 200 1000 1200,width=10cm
\includegraphics*[viewport=30 200 675 550,width=12cm,clip]{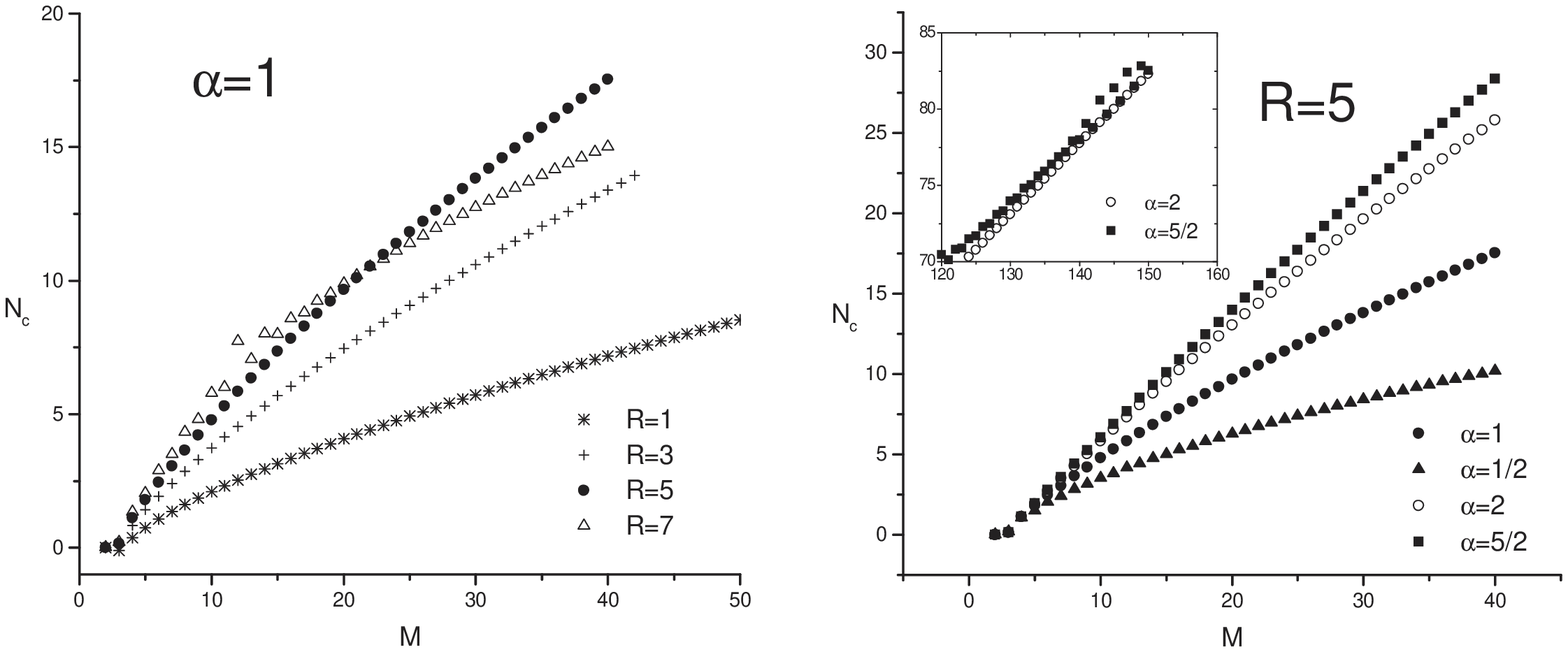}
\end{center}
\vskip-2cm
\caption{Illustration, on the
AO with $\protect\beta =1$, $m=2$, of the respective importance of the free
parameters $R$ and $\protect\alpha $, the correct values of which would be $%
R_{0}\simeq 5.192695$ and $\protect\alpha =2$. $N_{\mathrm{c}}$ is defined in the caption of figure \ref{fig4}. On the left, $\protect%
\alpha $ is fixed to $1$ and $R$ is varied. It is clear that the best
convergence corresponds to $R=5.$ For smaller values of $R$, the convergence seems
to be monotonically increasing whereas for $R=7$, it seems to be limited. On the right, $R$ is fixed to $5$ and $\protect\alpha $ is varied. The
best convergence seems to correspond to $\protect\alpha =5/2$ which is a too
large value. Actually, the small rule box shows that at larger values of the
order $M$, this choice, most probably, is no longer the best one.}
\label{fig6}
\end{figure}
%*************************************************************************

Figure \ref{fig1} shows that, presumably, all the zeroes of the fundamental
state $\psi _{0}\left( x\right) $ are located on the pure imaginary axis of
the $x$-plane. Hence, in the $z$-plane of the function $g\left( z\right) ,$
which is related to $\psi \left( x\right) $ via (\ref{eq:MansRicatti}, \ref%
{eq:gTaylor0}, \ref{eq:gTaylor}), those zeroes become singularities located
on the negative part of the real $z$-axis. The distance to the origin of the
closest singularity provides the value $R_{0}\simeq 5.192695$ and the plane
cut on the negative real axis, starting from the point $z=-5.192695,$ forms
an angular sector such that $R=R_{0}$ and $\alpha =\alpha _{0}=2$. Choosing
those values in the conformal mapping (\ref{eq:Map}, \ref{eq:MapInverse}),
produces a truncated series $\tilde{g}_{M}\left( w\right) $ in powers of $w$%
. As $M$ is increased the sum of $\tilde{g}_{M}\left( w\right) $ at $w=1$
should approach the asymptotic behavior corresponding to (\ref{eq:CondW}, %
\ref{eq:gTaylor0}, \ref{eq:gTaylor}). Thus, accounting for those
constraints, the auxiliary condition looked for to estimate $E_{0}$ may be
expressed as:%
\begin{equation}
\tilde{g}_{M}^{(2)}\left( w\right) _{w=1}=0  \label{eq:g2w1}
\end{equation}%
in which $\tilde{g}^{(2)}\left( w\right) $ stands for the conformal mapping
applied on the function $g^{\prime \prime }\left( z\right) $ which,
according to (\ref{eq:CondW}, \ref{eq:gTaylor0}, \ref{eq:gTaylor}), goes to
zero as $z\rightarrow \infty $.

The condition (\ref{eq:g2w1}) is a polynomial equation for $E$, the
solutions of which effectively display a stable real value as $M$ is
increased. Figure (\ref{fig5}) shows that, even when $M$ is small, this
value is easily identified and can be followed without difficulty. Actually
this stable value converges quickly to the true value $E_{0}.$ As example,
with $M=150$, $E_{0}$ is determined that way with an accuracy of $83$
significant figures (see figure \ref{fig4} and the appendix). Only the
choice of the order $M$ limits the accuracy of this estimation which is much
more accurate than the previously published values \cite{6352,6358,6339}.

An efficient and very convenient variant to the auxiliary condition (\ref%
{eq:g2w1}) consists in simply imposing the vanishing of the last term of the
series (\ref{eq:SerieAfterMap}), say 
\begin{equation}
u_{M}\left( E\right) =0\,.  \label{eq:practical}
\end{equation}

This condition avoids the explicit account of the asymptotic behavior of $%
\psi _{0}\left( x\right) $. That is quite justified knowing that the general
solution (for a generic $E$) has a moving singularity located right on the
positive real $w$-axis which generally limits the convergence of the series $%
\tilde{g}_{M}\left( w\right) $ except for the value $E_{0}$ for which this
moving singularity is sent to infinity. Hence imposing the condition (\ref%
{eq:practical}) amounts to force the convergence of the series and this
procedure enable the determination of the value of $E_{0}$.

\subsection{Practical use of the mapping method}

In general one does not know a priori the values $R_{0}$ and $\alpha _{0}$
which correspond to the location of the singularity the closest to the
origin of the solution $W_{0}\left( x\right) $ looked for. One must thus
consider $R$ and $\alpha $ as free parameters. Several procedures may be
conceived to approximately determine the "best" values of $R$ and $\alpha $
(which, ideally, should be close to $R_{0}$ and $\alpha _{0}$ respectively).
The following one is illustrated on the case $\beta =1$ and $m=2$.

Suppose first that $R_{0}$ and $\alpha _{0}$ define an angular sector in
which the solution looked for is analytic (i.e., there is no supplementary
singularity in the interior of this sector). Then, if $R<R_{0}$ and $\alpha
<\alpha _{0}$, the mapping method should provide a convergent result as the
order $M$ is increased. If it does not, then one of the previous inegalities
is not satisfied and a decrease of the trial value of $R$ (also of $\alpha $
even) is necessary. One thus proceeds by successive trial and error in order
to determine at least one couple of value $\left\{ R_{1},\alpha _{1}\right\} 
$ such that the mapping procedure begins to converge. In the example
presently chosen, if one sets a priori $R=1$ and $\alpha =1$, one already
gets, with $M=50$ an estimate of $E_{0}$ with 8 significant figures
(1.39235163). Having this first estimate, one may look at the effective
radius of convergence of the original series. This may be done with the help
of the d'Alembert or the Cauchy rule or (in the case where they do not
converge) via the plot of the sum of the series as function of $z$. This
provides an estimate of $R_{0}$ which may be used to get a better
convergence towards $E_{0}$. It is important to realize that, to get a
relatively large accuracy on $E_{0}$, it is not necessary to know $R_{0}$
with much accuracy.

The determination of the best value of $\alpha $ proceeds also by trial and
error and is determined by the criterium of best convergence as shown by
figure (\ref{fig6}). However, for small values of $M$, it may appears that
"exotic" effective best values of $\alpha $ may be observed. Figure (\ref%
{fig6}) shows that in the example of the AO presently treated, the value $%
\alpha =5/2$ provides, for small $M$, an apparent better convergence than
the right value $\alpha =2$. But, for larger values of $M$, one observes
that the former case finally yields perturbed convergence whereas with the
second case the convergence remains smooth [see figure (\ref{fig6})].

\subsection{Excited states}

%*************************************************************************
% Figure 7 - Convergence
\begin{figure}[tbp]
\begin{center} %viewport=200 200 1000 1200,width=10cm
\includegraphics*[viewport=-20 0 550 550,width=15cm,clip]{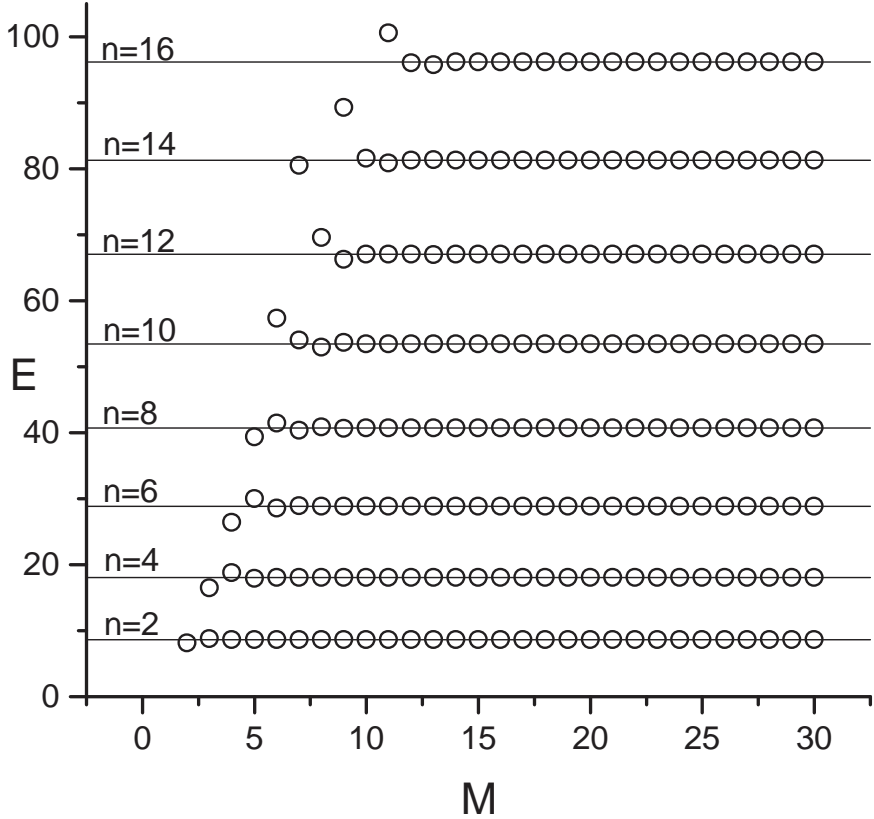}
\end{center}
\vskip-2cm
\caption{Evolution with $M$ of the
distribution of the real zeros (open circles) corresponding to the auxiliary
condition of the mapping method applied on the function $P\left( x\right) $
defined by (\protect\ref{eq:Factor}) in the case $\protect\beta =1$ and $m=2$
(even case). The horizontal lines correspond to the "exact" values of the
spectrum given in the appendix.}
\label{fig7}
\end{figure}
%*************************************************************************

The determination of the excited states proceeds as in section \ref{contour}%
. The fundamental state $W_{0}\left( x\right) $ (or the first odd state $%
W_{1}\left( x\right) $) is used as a basis through (\ref{eq:Factor}).
Consequently, there is no need to look for new determinations of $R$ and $%
\alpha $. The values obtained previously for $W_{0,1}\left( x\right) $ work
also for the excited states. At a given order $M$, the corresponding
auxiliary condition yields a polynomial equation for $E$, the real zeroes of
which form, once and for all, the approximate spectrum of the AO which can
be, step by step, unambiguously determined as shown by figure (\ref{fig7}).
The values of the spectrum are given in the appendix, they are much more
accurate than (and compatible with) the existing previous estimates \cite%
{6352,6358,6339}.

\subsection{Other examples treated}

In this part configurations which are more difficult to treat than the
previous one are considered. One easily understands that the larger $m$, the
harder the determination of the spectrum. In principle the difficulty also
grows as $\beta $ decreases: for a fixed $m$, the limit $\beta \rightarrow 0$
is the limit of infinite coupling constant $\lambda \rightarrow \infty $.
Another kind of complication arises when $\beta $ is more and more negative
(double-well configurations): a Taylor-series-based method is not very
appropriate because the expansion is performed on the top of the hill of the
potential.

\subsubsection{The AO with $\protect\beta =0$ and $m=2$}

This is the example treated explicitly in \cite{6394} where, with $M=300$,
the fundamental energy $E_{0}$ had been determined with an accuracy of $65$
significant figures (48 figures for the excited-state energies $E_{n}$ up to 
$n=39$).

The analytic structure of $W_{0}\left( x\right) $ is similar to that of the
case $\beta =1$ previously considered: all the singularities are probably
located on the imaginary axis of the complex $x$-plane. The location of the
singularity the closest to the origin implies $R_{0}\simeq 4.2526611666886$
whereas $\alpha _{0}=2$.

Using those values to fix $R$ and $\alpha $, the mapping method yields, for $%
M=150$, an estimate of $E_{0}$ with already $76$ significant figures.
Moreover, using the rescaling of \cite{6394,6398} and a series calculated up
to the order $M=250$, the number of significant figures obtained is $114.$
The accuracy is even better in the odd case since, at the same order $M=250$%
, $E_{1}$ is estimated with an accuracy of $120$ significant figures (see
the appendix).

As for the excited states both in the even and odd cases and $M=180$, the 40
first energy states are estimated with an accuracy larger than $56$ figures
in the even case (63 in the odd case).

All these estimates are presented in the appendix. They all coincide with
the estimates given in \cite{6394} but, again, involve more significant
figures.

\subsubsection{The AO with $\protect\beta =0$ and $m=3$}

In that example, the singularity of $W_{0}\left( x\right) $ are not located
on the imaginary axis of the $x$-plane but on four straight lines starting
from the origin so that the singularities the closest to the origin are
located at points $re^{\pm i\theta }$ and $re^{\pm i\left( \pi -\theta
\right) }$ with:%
\begin{eqnarray*}
r &\simeq &1.81997592362 \\
\theta &\simeq &0.3663025\times \pi
\end{eqnarray*}

Considering now the complex $z$-plane, and the generic definition of $R$ and 
$\alpha $ associated with an angular sector as drawn on figure (\ref{fig3}),
this induces the angular sector defined by:%
\begin{eqnarray*}
R_{0} &\simeq &3.31231236256 \\
\alpha _{0} &\simeq &0.732605
\end{eqnarray*}%
with two singularities located on its edges. With such a sector of
analyticity, without singularity taking place on the negative side of the
real $z$-axis, the parameter $R$ is no longer bound by the condition $R\leq
R_{0}$ and may be chosen as large as one wants provided that $\alpha $ be
modified adequately. However the larger $R,$ the smaller $\alpha $, and it
appears that it is preferable to choose the largest integer value of $\alpha 
$ which is compatible with the constraint $R>0$. Presently the value $\alpha
=1$ may be chosen whith $R\simeq 2.2107.$ With those values and the
rescaling of \cite{6394,6398}, using $M=250$, the fundamental energy $E_{0}$
has been determined with 56 significant figures. The 20 first even excited
states are then determined with at least 24 significant figures with $M=181.$
Similar accuracies are obtained in the odd case (see the appendix).

\subsubsection{Double well\label{double}}

The more $\beta $ is negative, the more the calculations based on a Taylor
expansion about the origin $x=0$ are difficult. In addition, one will see
that the radius of convergence of the Taylor series for $W_{0,1}\left(
x\right) $ decreases so as to considerably reduce the efficiency of the
mapping method (although it continues to work correctly). For small $\beta
<0 $ and $m=2$, the mapping method is still very efficient as shown by the
estimates given in the appendix for $\beta =-1$ and $m=2$.

For the sake of an illustration of the difficulties encountered for
increasingly negative values of $\beta $, the following is limited to a
short presentation of the results for two configurations: $\beta =-11,$ $m=3$
and $\beta =-10,$ $m=2$ (the configuration $\beta =-50,$ $m=2$ is also
shortly alluded to).

\paragraph{$\protect\beta =-11,$ $m=3$}

According to section \ref{Quasi}, that configuration is quasi-exact. One
finds that the singularities of $W_{0}\left( x\right) $ are all located on
the imaginary axis in the complex $x$-plane and the optimal values of the
mapping parameters are determined to be:%
\begin{eqnarray*}
R_{0} &\simeq &0.29289 \\
\alpha _{0} &=&2
\end{eqnarray*}

According to (\ref{eq:CondW}), the auxiliary condition is:%
\begin{equation}
\tilde{g}_{M}^{(2)}\left( w\right) _{w=1}=-\frac{1}{2}
\end{equation}%
With $M=100$, the estimate obtained with the mapping method for the
fundamental-state energy is:%
\begin{equation*}
E_{0}=-7.99999999999999969\pm 1.4\times 10^{-15}
\end{equation*}%
which corresponds to the first exact value $-8$ listed in section \ref{Quasi}%
.

In the odd case and $M=100$ also, $E_{1}$ (which is not an \textquotedblleft
exact\textquotedblright\ value) is better determined with an accuracy of $30$
significant figures, using $R_{0}\simeq 0.9129007$ and $\alpha _{0}=2.$ It
comes:%
\begin{equation*}
E_{1}=-7.91735504373511193638182033627
\end{equation*}

To get the two other exact values, one must consider the even excited
states. With $M=100$, the results are:%
\begin{eqnarray*}
E_{2} &=&0.00004\pm 0.0007 \\
E_{4} &=&7.99976\pm 0.00397
\end{eqnarray*}%
which correspond to the exact values $0$ and $8$ of section \ref{Quasi}
respectively.

The two following odd excited states are again better determined:

\begin{eqnarray*}
E_{3} &=&2.520358324953 \\
E_{5} &=&14.11296433804
\end{eqnarray*}

To improve the accuracy, one should have to consider larger values of $M$.

One notices that, compared to the even case, the better accuracy obtained in
the odd case coincides with a larger radius $R_{0}$. The following example
confirms this (expected) correlation.

\paragraph{$\protect\beta =-10,$ $m=2$}

The singularities of $W_{0}\left( x\right) $ are again all located on the
imaginary axis in the complex $x$-plane and the optimal values of the
mapping parameters are found to be:%
\begin{eqnarray*}
R_{0} &\simeq &0.118685154 \\
\alpha _{0} &=&2
\end{eqnarray*}

According to (\ref{eq:CondW}), the auxiliary condition is:%
\begin{equation}
\tilde{g}_{M}^{(2)}\left( w\right) _{w=1}=0
\end{equation}

The radius $R_{0}$ is smaller than in the previous case and a less accurate
estimate must be expected. Effectively with $M=126$, the estimate obtained
with the mapping method for the fundamental-state energy displays only 9
significant figures:%
\begin{equation*}
E_{0}=-20.63357672
\end{equation*}%
and for $M=200$, one gets (after a relatively long time of calculation)
about 16 significant figures only:%
\begin{equation*}
E_{0}=-20.633576702947807
\end{equation*}

In the odd case, one finds:%
\begin{eqnarray*}
R_{0} &\simeq &0.44986354 \\
\alpha _{0} &=&2
\end{eqnarray*}%
and the estimate with $M=100$ is found with an accuracy of 16 significant
figures:%
\begin{equation*}
E_{1}=-20.63354688440494
\end{equation*}

The above estimates of $E_{0}$ and $E_{1}$ are not as accurate as those
obtained using the Pad\'{e} method with $M\simeq 40$ only \cite{6389}:%
\begin{eqnarray*}
E_{0} &=&-20.633576702947799150 \\
E_{1} &=&-20.633546884404911079
\end{eqnarray*}

One sees that, in accordance with its nature, the Pad\'{e} method makes the
best with the (isolated) singularities located on the negative real $z$-axis
even if one of them comes close to the origin. On the contrary, the
efficiency of the mapping method decreases in that case because it is
intimately tied to the location of the singularity the closest to the origin.

\paragraph{$\protect\beta =-50,$ $m=2$}

It is worth mentionning that, in that case, the radius $R_{0}$ is found to
be close to $0.004012$. With a so small value, the mapping method is
practically useless (although it continues to be valid in principle but
large values of $M$ are already required to simply get an idea on the order
of magnitude of $E_{0}$). In the circumstances the analytic-continuation
method of section \ref{Continuation} is particularly useful since, as
already mentionned, that method yields an estimate of $E_{0}$ with $22$
significant figures \cite{6431} with $M\simeq 40$ only.

\section{Summary and conclusion\label{Conc}}

The AO with different potential configurations have been explicitly
considered to illustrate the efficiency and the limitations of a
quasi-analytic method for solving nonlinear ODE named the mapping method 
\cite{6319}. As several other quasi-analytic methods encountered here and
there in the literature of the AO it is based on a generic Taylor series in
powers of the independent variable. After a short presentation of those
methods, the mapping method has been introduced and its use clearly
illustrated on the basis of the simplicity of the analytic properties of the
AO. In particular the practical determination of the two adjustable
parameters $R$ and $\alpha $, inherent to the method [see eqs. (\ref{eq:Map}%
, \ref{eq:MapInverse})], has been shown to correspond precisely to those
analytic properties. In absence of any information on them, the criterion of
best convergence is appeared to be valid. Extremely high accurate estimates
of the spectrum of the AO have been easily obtained (see the appendix)
provided that the effective value of $R$ was not \textquotedblleft
too\textquotedblright\ small since, in accordance with its nature, the
efficiency of the method decreseases with $R$. In the case of very small $R$%
, the analytic-continuation method or the Pad\'{e} method have appeared to
be more efficient. Finally the mapping method \cite{6319} is an extremely
refined method, easy to use and which provides clear convergences towards
the values looked for provided the radius of convergence of the initial
Taylor series is not too small. The Pad\'{e} method \cite{6201} is also an
easy-to-use method which is more robust but less refined than the mapping
method. The two methods (and also the analytic-continuation method \cite%
{6449,6432}) may certainly be advantageously associated in the process of
solving a two-point boundary problem of a nonlinear ODE.

\ack I am indebted to Bruno Boisseau and Hector Giacomini for numerous
discussions and encouragements all along this work.

\appendix

\section*{Appendix}

\setcounter{section}{1}

In this appendix are presented the estimates of the energy spectrum of
several configurations of the AO as obtained using the mapping method.

\subsection{Spectrum for $\protect\beta =1,$ $m=2$}

With $M=150$, $R=5.1926948467766235669$ and $\alpha =2$, the fundamental
energy state has been determined with 83 significant figures:%
\begin{eqnarray*}
E_{0}=1. &&39235164153029185565750787660993418460006 \\
&&67112208340889063493238775674318756465286
\end{eqnarray*}

The estimates of the 17 first even excited states, obtained with $M=115$,
are given in table \ref{Tab1}.

\begin{table}[tbp]
\centering
\par
\begin{tabular}{ccc}
\hline\hline
$n$ & $E_{n}$ & $N_{c}$ \\ \hline
$2$ & \multicolumn{1}{l}{$8.655049957759309688116539457377308026275$} & $40$
\\ 
$4$ & \multicolumn{1}{l}{$18.0575574363032528947712396465254348531$} & $39$
\\ 
$6$ & \multicolumn{1}{l}{$28.83533845950424884013363571549983817$} & $37$ \\ 
$8$ & \multicolumn{1}{l}{$40.690386082106444725278931481582464$} & $35$ \\ 
$10$ & \multicolumn{1}{l}{$53.4491021396652646008315064597595$} & $33$ \\ 
$12$ & \multicolumn{1}{l}{$66.995030001247166061019704904702$} & $32$ \\ 
$14$ & \multicolumn{1}{l}{$81.24350505076715273706652147034$} & $31$ \\ 
$16$ & \multicolumn{1}{l}{$96.129642045234052046811222396$} & $29$ \\ 
$18$ & \multicolumn{1}{l}{$111.6018150451729585337015116$} & $28$ \\ 
$20$ & \multicolumn{1}{l}{$127.617777795354918333962292$} & $27$ \\ 
$22$ & \multicolumn{1}{l}{$144.142195296398163731983$} & $24$ \\ 
$24$ & \multicolumn{1}{l}{$161.14499069451295186862$} & $23$ \\ 
$26$ & \multicolumn{1}{l}{$178.6001923668757611938$} & $22$ \\ 
$28$ & \multicolumn{1}{l}{$196.48510291022044366$} & $20$ \\ 
$30$ & \multicolumn{1}{l}{$214.779683549176627$} & $18$ \\ 
$32$ & \multicolumn{1}{l}{$233.4660874793752$} & $16$ \\ 
$34$ & \multicolumn{1}{l}{$252.5282990614935$} & $16$ \\ \hline
\end{tabular}%
\caption{Estimates of even excited state energies fo $\protect\beta =1,$ $%
m=2 $ as obtained by the mapping method with $M=115$. The last column
displays an approximate value of the number of significant figures obtained
in each case.}
\label{Tab1}
\end{table}

With $M=115$, $R=6.03344983950017$ and $\alpha =2$ the first odd state has
been determined with 70 significant figures:%
\begin{eqnarray*}
E_{1}=4. &&64881270421207753637703291726058448 \\
&&8898860447882825934823424910341006
\end{eqnarray*}

The estimates of the 20 first odd excited states, obtained with $M=121$, are
given in table \ref{Tab2}.

\begin{table}[tbp]
\centering
\par
\begin{tabular}{ccc}
\hline\hline
$n$ & $E_{n}$ & $N_{c}$ \\ \hline
$3$ & \multicolumn{1}{l}{$13.1568038980498750792097720403823146746501484$} & 
$45$ \\ 
$5$ & \multicolumn{1}{l}{$23.2974414512231890848644819920981238281208$} & $%
42 $ \\ 
$7$ & \multicolumn{1}{l}{$34.640848321111332542884527618156342033769$} & $41$
\\ 
$9$ & \multicolumn{1}{l}{$46.9650095056755279840964433241751142524$} & $39$
\\ 
$11$ & \multicolumn{1}{l}{$60.12952295915777131584801605985282216$} & $37$
\\ 
$13$ & \multicolumn{1}{l}{$74.03587435910253018074120548740369915$} & $37$
\\ 
$15$ & \multicolumn{1}{l}{$88.61034880079915887303910537132488$} & $34$ \\ 
$17$ & \multicolumn{1}{l}{$103.7953003222726096781116879557136$} & $34$ \\ 
$19$ & \multicolumn{1}{l}{$119.54417073305031113002694934564$} & $32$ \\ 
$21$ & \multicolumn{1}{l}{$135.818417325610373340451430114$} & $30$ \\ 
$23$ & \multicolumn{1}{l}{$152.5855042055739215668661903$} & $28$ \\ 
$25$ & \multicolumn{1}{l}{$169.817528001595348199877321$} & $27$ \\ 
$27$ & \multicolumn{1}{l}{$187.4902426929503225448058$} & $25$ \\ 
$29$ & \multicolumn{1}{l}{$205.58234660442351871834$} & $23$ \\ 
$31$ & \multicolumn{1}{l}{$224.0749478526003062853$} & $22$ \\ 
$33$ & \multicolumn{1}{l}{$242.95115495114712353$} & $20$ \\ 
$35$ & \multicolumn{1}{l}{$262.1957574685198472$} & $19$ \\ 
$37$ & \multicolumn{1}{l}{$281.794972923819312$} & $18$ \\ 
$39$ & \multicolumn{1}{l}{$322.008069744848$} & $15$ \\ 
$41$ & \multicolumn{1}{l}{$363.5018948643$} & $13$ \\ \hline
\end{tabular}%
\caption{Estimates of odd excited state energies fo $\protect\beta =1,$ $m=2$
as obtained by the mapping method with $M=121$. The last column displays an
approximate value of the number of significant figures obtained in each
case. }
\label{Tab2}
\end{table}

All the estimates obtained are in agreement with the existing literature 
\cite{6348,6339,6352} at least up to 10--16 significant figures they quote.

\subsection{Spectrum for $\protect\beta =0,$ $m=2$}

In order to compare the efficiency of the respective methods, the rescaling
of \cite{6392} has been used in the study of this configuration. In the even
case, the parameters of the conformal mapping have been fixed to $%
R=0.7515601141384727$ and $\alpha =2$ corresponding to the location of the
singularities of $W_{0}\left( x\right) $ in the complex $x$-plane. With $%
M=250$, the fundamental state energy has been determined with 114
significant figures:%
\begin{eqnarray*}
E_{0}=1. &&06036209048418289964704601669266354551 \\
&&52087285289779332162452416959435630443 \\
&&4442112689629913467170351054624435858
\end{eqnarray*}

The 20 first even excited state, with $M=180$, have been determined with an
accuracy as displayed in table \ref{Tab3}.

\begin{table}[tbp]
\centering
\par
\begin{tabular}{ccc}
\hline\hline
$n$ & $E_{n}$ & $N_{c}$ \\ \hline
$2$ & \multicolumn{1}{l}{$%
\begin{array}{c}
7.455697937986738392156591347185767488137819536 \\ 
\multicolumn{1}{r}{74910086643688870645553641487360130430909}%
\end{array}%
$} & $86$ \\ 
$4$ & \multicolumn{1}{l}{$%
\begin{array}{c}
16.26182601885022593789495443038461353424458650 \\ 
\multicolumn{1}{r}{45141645421211209355815531735099805375715}%
\end{array}%
$} & $86$ \\ 
$6$ & \multicolumn{1}{l}{$%
\begin{array}{c}
26.52847118368251819181382818368122792897571351 \\ 
\multicolumn{1}{r}{294412586757107839537851365173102739202}%
\end{array}%
$} & $84$ \\ 
$8$ & \multicolumn{1}{l}{$%
\begin{array}{c}
37.92300102703398514651637855190995445547624176 \\ 
\multicolumn{1}{r}{6768540940158056680472093643981051811}%
\end{array}%
$} & $82$ \\ 
$10$ & \multicolumn{1}{l}{$%
\begin{array}{c}
50.25625451668291903974458810526275993610124418 \\ 
\multicolumn{1}{r}{075701487761224602795878984785502928}%
\end{array}%
$} & $81$ \\ 
$12$ & \multicolumn{1}{l}{$%
\begin{array}{c}
63.40304698671889497209348492294463561729801089 \\ 
\multicolumn{1}{r}{28673988084514263379591656250007033}%
\end{array}%
$} & $80$ \\ 
$14$ & \multicolumn{1}{l}{$%
\begin{array}{c}
77.27320048198396570945437444123301297760205161 \\ 
\multicolumn{1}{r}{696448894060040064817405480052452}%
\end{array}%
$} & $78$ \\ 
$16$ & \multicolumn{1}{l}{$%
\begin{array}{c}
91.79806680899113406984882666252278502713173398 \\ 
\multicolumn{1}{r}{90657479115965604376782712914156}%
\end{array}%
$} & $77$ \\ 
$18$ & \multicolumn{1}{l}{$%
\begin{array}{c}
106.9233073817325256530751021361667347590846183 \\ 
\multicolumn{1}{r}{2008357932027830540845145707859}%
\end{array}%
$} & $76$ \\ 
$20$ & \multicolumn{1}{l}{$%
\begin{array}{c}
122.6046390009994550207629714176151818749766332 \\ 
\multicolumn{1}{r}{22893621015463272615565590626}%
\end{array}%
$} & $74$ \\ 
$22$ & \multicolumn{1}{l}{$%
\begin{array}{c}
138.8051479113947530940883610374380934368840381 \\ 
\multicolumn{1}{r}{671106897341615306983140020}%
\end{array}%
$} & $72$ \\ 
$24$ & \multicolumn{1}{l}{$%
\begin{array}{c}
155.4935022686820551465506277045177934315466954 \\ 
\multicolumn{1}{r}{5414662134066104062468041}%
\end{array}%
$} & $70$ \\ 
$26$ & \multicolumn{1}{l}{$%
\begin{array}{c}
172.6427119628444830668784007249101646087468190 \\ 
\multicolumn{1}{r}{46656519547351087844600}%
\end{array}%
$} & $68$ \\ 
$28$ & \multicolumn{1}{l}{$%
\begin{array}{c}
190.2292386524627088279661423652564958817125565 \\ 
\multicolumn{1}{r}{189811610764662196522}%
\end{array}%
$} & $66$ \\ 
$30$ & \multicolumn{1}{l}{$%
\begin{array}{c}
208.2323390051439485027234771412174093538259896 \\ 
\multicolumn{1}{r}{12647599436328976471}%
\end{array}%
$} & $65$ \\ 
$32$ & \multicolumn{1}{l}{$%
\begin{array}{c}
226.6335684811375306036977950956875757647131141 \\ 
\multicolumn{1}{r}{4054544586360420784}%
\end{array}%
$} & $64$ \\ 
$34$ & \multicolumn{1}{l}{$%
\begin{array}{c}
245.4163987919363679916747517301130143224691180 \\ 
\multicolumn{1}{r}{4034148423591243}%
\end{array}%
$} & $61$ \\ 
$36$ & \multicolumn{1}{l}{$%
\begin{array}{c}
264.5659178144988804976115801604901826931113435 \\ 
\multicolumn{1}{r}{585310082273247}%
\end{array}%
$} & $60$ \\ 
$38$ & \multicolumn{1}{l}{$%
\begin{array}{c}
284.0685905814007431504962812081250647770847132 \\ 
\multicolumn{1}{r}{6709967957974}%
\end{array}%
$} & $58$ \\ 
$40$ & \multicolumn{1}{l}{$%
\begin{array}{c}
303.9120663483839986502856546215441904999429332 \\ 
\multicolumn{1}{r}{70794969657}%
\end{array}%
$} & $56$ \\ \hline\hline
\end{tabular}%
\caption{Estimates of even excited state energies fo $\protect\beta =0,$ $%
m=2 $ as obtained by the mapping method with $M=180$. For each $n$ a second
line displays the continuation of digits. The last column displays an
approximate value of the number of significant figures obtained in each
case. }
\label{Tab3}
\end{table}

In the odd case, the mapping parameters are fixed to $R=0.65325380587388$
and $\alpha =2$. The first odd energy state has been determined with $120$
significant figures ($M=250$):%
\begin{eqnarray*}
E_{1}=3. &&7996730298013941687830941885125689577660 \\
&&6546732735340998309740435228060023769937 \\
&&931149834764800025936072118065726127409
\end{eqnarray*}

With $M=181$, the 20 first odd excited states are estimated as shown in
table \ref{Tab4}.

\begin{table}[tbp]
\centering
\par
\begin{tabular}{ccc}
\hline\hline
$n$ & $E_{n}$ & $N_{c}$ \\ \hline
3 & \multicolumn{1}{l}{$%
\begin{array}{c}
11.64474551137816202085037328137093643655087216197 \\ 
\multicolumn{1}{r}{4453001669270913851625610586869623215876947849}%
\end{array}%
$} & 94 \\ 
5 & \multicolumn{1}{l}{$%
\begin{array}{c}
21.23837291823594002414971111358863637670483205965 \\ 
\multicolumn{1}{r}{52359735931435198551551347510844880262694475}%
\end{array}%
$} & 92 \\ 
7 & \multicolumn{1}{l}{$%
\begin{array}{c}
32.09859771096832663427210643833189668755992423305 \\ 
\multicolumn{1}{r}{978762342171173242882982430096400348694857}%
\end{array}%
$} & 90 \\ 
9 & \multicolumn{1}{r}{$%
\begin{array}{c}
43.98115809728973078531811375282736479477947632092 \\ 
\multicolumn{1}{r}{7591678745311041539444158927512411142579}%
\end{array}%
$} & 88 \\ 
11 & \multicolumn{1}{r}{$%
\begin{array}{c}
56.73421405517303604730167103345175751218649581511 \\ 
\multicolumn{1}{r}{39323399711320613883016063024691753961}%
\end{array}%
$} & 86 \\ 
13 & \multicolumn{1}{r}{$%
\begin{array}{c}
70.25239462861659088913085193801913436712112961025 \\ 
\multicolumn{1}{r}{6667440126160413192685489535915017187}%
\end{array}%
$} & 85 \\ 
15 & \multicolumn{1}{l}{$%
\begin{array}{c}
84.45746627494195892888833727655962244102394532902 \\ 
\multicolumn{1}{r}{36232438363406557704202821636803725}%
\end{array}%
$} & 83 \\ 
17 & \multicolumn{1}{l}{$%
\begin{array}{c}
99.28860666049327964581980672652356738129618429531 \\ 
\multicolumn{1}{r}{5439623923490779365455043123425144}%
\end{array}%
$} & 82 \\ 
19 & \multicolumn{1}{l}{$%
\begin{array}{c}
114.6969173849851755381509283197832144905718574275 \\ 
\multicolumn{1}{r}{094513375030013226382190538593688}%
\end{array}%
$} & 81 \\ 
21 & \multicolumn{1}{l}{$%
\begin{array}{c}
130.6420687486299788272084813607924461800377164913 \\ 
\multicolumn{1}{r}{39742213272421838252016675620413}%
\end{array}%
$} & 80 \\ 
23 & \multicolumn{1}{l}{$%
\begin{array}{c}
147.0901212576038250593001573616393276076304105930 \\ 
\multicolumn{1}{r}{039522847501664310955551882108}%
\end{array}%
$} & 78 \\ 
25 & \multicolumn{1}{l}{$%
\begin{array}{c}
164.0120436228653346009732131640724071039525335388 \\ 
\multicolumn{1}{r}{6631868425810048601661780456}%
\end{array}%
$} & 76 \\ 
27 & \multicolumn{1}{l}{$%
\begin{array}{c}
181.3826661857682257850039425558429144469164137642 \\ 
\multicolumn{1}{r}{86635610868225155373630509}%
\end{array}%
$} & 74 \\ 
29 & \multicolumn{1}{l}{$%
\begin{array}{c}
199.1799188337470723626128506406160415222128605618 \\ 
\multicolumn{1}{r}{585703827731174998592784}%
\end{array}%
$} & 72 \\ 
31 & \multicolumn{1}{l}{$%
\begin{array}{c}
217.3842616741034519777291998578486546497746509056 \\ 
\multicolumn{1}{r}{71607931000643490074315}%
\end{array}%
$} & 71 \\ 
33 & \multicolumn{1}{l}{$%
\begin{array}{c}
235.9782503616957543907796533463549468434869938281 \\ 
\multicolumn{1}{r}{882674294457757985404}%
\end{array}%
$} & 69 \\ 
35 & \multicolumn{1}{l}{$%
\begin{array}{c}
254.9461979707982589533328548019610045320736538738 \\ 
\multicolumn{1}{r}{26605804407445584226}%
\end{array}%
$} & 68 \\ 
37 & \multicolumn{1}{l}{$%
\begin{array}{c}
274.2739076589405355230240455122439840737715234577 \\ 
\multicolumn{1}{r}{772581092942928166}%
\end{array}%
$} & 66 \\ 
39 & \multicolumn{1}{l}{$%
\begin{array}{c}
293.9484582660060854336699974835216263034458992746 \\ 
\multicolumn{1}{r}{8022674262238544}%
\end{array}%
$} & 64 \\ 
41 & \multicolumn{1}{l}{$%
\begin{array}{c}
313.9580301839779322009488360107644470523106026887 \\ 
\multicolumn{1}{r}{398971572795076}%
\end{array}%
$} & 63 \\ \hline\hline
\end{tabular}%
\caption{Estimates of odd excited state energies fo $\protect\beta =0,$ $m=2$
as obtained by the mapping method with $M=181$. For each $n$ a second line
displays the continuation of digits. The last column displays an approximate
value of the number of significant figures obtained in each case. }
\label{Tab4}
\end{table}

All the above values are in agreement with the estimates published by
Leonard and Mansfiled \cite{6394} at least up to 48--65 significant figures
they quote.

\subsection{Spectrum for $\protect\beta =0,$ $m=3$}

In this case the singularities of $W_{0}\left( x\right) $ are not located on
the imaginary axis of the complex $x$-plane. Hence $\alpha \neq 2$, and $R$
may be adjusted so that one may choose $\alpha =1$ to make the calculations
easier. Using the rescaling trick of \cite{6392}, for $R=0.42181$ and $M=250$
the fundamental energy has been determined with$\ 56$ significant figures:%
\begin{equation*}
E_{0}=1.1448024537970527637654575341495490765378125289798777292
\end{equation*}

With $M=181$ the 20 first excited even energy states are determined to be as
in table \ref{Tab5}.

\begin{table}[tbp]
\centering
\par
\begin{tabular}{ccc}
\hline\hline
$n$ & $E_{n}$ & $N_{c}$ \\ \hline
$2$ & \multicolumn{1}{l}{$9.073084560921433856016249096664712179552456747$}
& $46$ \\ 
$4$ & \multicolumn{1}{l}{$21.7141654221967222816897851486065849098555916$} & 
$45$ \\ 
$6$ & \multicolumn{1}{l}{$37.6130865608951608896015430474867021943435$} & $%
42 $ \\ 
$8$ & \multicolumn{1}{l}{$56.199300852499359421090060527088020427952$} & $41$
\\ 
$10$ & \multicolumn{1}{l}{$77.12734146382543798734243263215304574937$} & $40$
\\ 
$12$ & \multicolumn{1}{l}{$100.15892789265287617620112949318711203$} & $38$
\\ 
$14$ & \multicolumn{1}{l}{$125.11668074758135609314911179285667525$} & $38$
\\ 
$16$ & \multicolumn{1}{l}{$151.86183175856232716716355175727249217$} & $38$
\\ 
$18$ & \multicolumn{1}{l}{$180.28193894068557040564022178012003$} & $35$ \\ 
$20$ & \multicolumn{1}{l}{$210.2834723116004980559950901608572$} & $34$ \\ 
$22$ & \multicolumn{1}{l}{$241.787034056498822149736098165624$} & $33$ \\ 
$24$ & \multicolumn{1}{l}{$274.72411811463802668627165975505$} & $32$ \\ 
$26$ & \multicolumn{1}{l}{$309.03482436375835367061889961183$} & $32$ \\ 
$28$ & \multicolumn{1}{l}{$344.666193379464295659578007932$} & $30$ \\ 
$30$ & \multicolumn{1}{l}{$381.57096047135932038713367165$} & $29$ \\ 
$32$ & \multicolumn{1}{l}{$419.7066022261534757147058497$} & $28$ \\ 
$34$ & \multicolumn{1}{l}{$459.034592731153881282354894$} & $27$ \\ 
$36$ & \multicolumn{1}{l}{$499.5198136445789902188947$} & $25$ \\ 
$38$ & \multicolumn{1}{l}{$541.130079449256802989435$} & $24$ \\ 
$40$ & \multicolumn{1}{l}{$583.835750482220100204957$} & $24$ \\ \hline
\end{tabular}%
\caption{Estimates of even excited state energies fo $\protect\beta =0,$ $%
m=3 $ as obtained by the mapping method with $M=181$. For each $n$ a second
line displays the continuation of digits. The last column displays an
approximate value of the number of significant figures obtained in each
case. }
\label{Tab5}
\end{table}

A similar singularity distribution occurs for the odd state so that $\alpha $
may be fixed to 1 with $R=0.3467407074263.$ With $M=250$ and the rescaling
trick of \cite{6392}, the first odd energy state is determined with 59
significant figures:%
\begin{equation*}
E_{1}=4.3385987115139811916473368880601867793285474529432549486123
\end{equation*}

With $M=182$ the 20 first excited odd energy states are determined to be as
in table \ref{Tab6}.

\begin{table}[tbp]
\centering
\par
\begin{tabular}{ccc}
\hline\hline
$n$ & $E_{n}$ & $N_{c}$ \\ \hline
$3$ & \multicolumn{1}{l}{$14.93516963491073601571195495127066852852812677095$%
} & $49$ \\ 
$5$ & \multicolumn{1}{l}{$29.29964593740189312978407064699053036251622441$}
& $46$ \\ 
$7$ & \multicolumn{1}{l}{$46.59521144855173401972883161023883550190357379$}
& $46$ \\ 
$9$ & \multicolumn{1}{l}{$66.387281706591605561901323787042869206296808$} & $%
44$ \\ 
$11$ & \multicolumn{1}{l}{$88.392375769030215606669098836449373556533$} & $%
41 $ \\ 
$13$ & \multicolumn{1}{l}{$112.40643581144162471251824362883088085074$} & $%
41 $ \\ 
$15$ & \multicolumn{1}{l}{$138.2733726559526082048335446213274433132$} & $40$
\\ 
$17$ & \multicolumn{1}{l}{$165.86873924695253971482736881612442986$} & $38$
\\ 
$19$ & \multicolumn{1}{l}{$195.0902772336198933064947454865236007$} & $37$
\\ 
$21$ & \multicolumn{1}{l}{$225.852006093743593994689769811061848$} & $36$ \\ 
$23$ & \multicolumn{1}{l}{$258.08030930413349189359807845604983$} & $35$ \\ 
$25$ & \multicolumn{1}{l}{$291.7112248219607946631497420954846$} & $34$ \\ 
$27$ & \multicolumn{1}{l}{$326.688501175539696874664982050866$} & $33$ \\ 
$29$ & \multicolumn{1}{l}{$362.96216142162740255082506466132$} & $32$ \\ 
$31$ & \multicolumn{1}{l}{$400.4874160213062718461371957159$} & $31$ \\ 
$33$ & \multicolumn{1}{l}{$439.223822596901596341934330594$} & $30$ \\ 
$35$ & \multicolumn{1}{l}{$479.13462481132026473711629662$} & $29$ \\ 
$37$ & \multicolumn{1}{l}{$520.186224052830671932702626$} & $27$ \\ 
$39$ & \multicolumn{1}{l}{$562.347751461637717237705920$} & $27$ \\ 
$41$ & \multicolumn{1}{l}{$605.59071704038135107740483$} & $26$ \\ \hline
\end{tabular}%
\caption{Estimates of odd excited state energies fo $\protect\beta =0,$ $m=3$
as obtained by the mapping method with $M=182$. For each $n$ a second line
displays the continuation of digits. The last column displays an approximate
value of the number of significant figures obtained in each case. }
\label{Tab6}
\end{table}

All the above values are in agreement with the estimates published in \cite%
{6339} at least up to 16 significant figures quoted.

\subsection{Spectrum for $\protect\beta =-1,$ $m=2$}

In this case the zeroes of the two first states (even and odd) are located
on the imaginary axis of the complex $x$-plane and the parameter $\alpha $
of the mapping transform (\ref{eq:Map}) is set to 2. The values of $R$
corresponding to the location of the zeroes the closest to the origin in the
two cases are found to be:%
\begin{eqnarray*}
R &=&3.33794474927577803\qquad (\mathrm{even)} \\
R &=&4.31581775651355421\qquad (\mathrm{odd)}
\end{eqnarray*}

With $M=100$, the tables \ref{Tab7} and \ref{Tab8} give the values of the 21
first energy states in each cases (even and odd):

\begin{table}[tbp]
\centering
\par
\begin{tabular}{ccc}
\hline\hline
$n$ & $E_{n}$ & $N_{c}$ \\ \hline
$0$ & \multicolumn{1}{l}{$0.65765300518071512305902172311059356037493794193$}
& $47$ \\ 
$2$ & \multicolumn{1}{l}{$6.163901256963068240915237315023209615538385197555$%
} & $48$ \\ 
$4$ & \multicolumn{1}{l}{$14.3724065046778689559065676196721576344668654764$}
& $47$ \\ 
$6$ & \multicolumn{1}{l}{$24.12807549278232749310320568101826099342879528$}
& $45$ \\ 
$8$ & \multicolumn{1}{l}{$35.0621490310767668384022166011110081686251495$} & 
$44$ \\ 
$10$ & \multicolumn{1}{l}{$46.96996815990174028776218699362301862009641$} & $%
42$ \\ 
$12$ & \multicolumn{1}{l}{$59.71764010164776255353662335294352670074898$} & $%
42$ \\ 
$14$ & \multicolumn{1}{l}{$73.2094806376508714565578227848019199478959$} & $%
41$ \\ 
$16$ & \multicolumn{1}{l}{$87.3730816075075214515884652935289486513$} & $38$
\\ 
$18$ & \multicolumn{1}{l}{$102.151393199174403996826472776949609559$} & $38$
\\ 
$20$ & \multicolumn{1}{l}{$117.49809600941412545762369437004927502$} & $37$
\\ 
$22$ & \multicolumn{1}{l}{$133.37469793993799359588507824136165$} & $34$ \\ 
$24$ & \multicolumn{1}{l}{$149.7486123997828373538531290781556$} & $33$ \\ 
$26$ & \multicolumn{1}{l}{$166.59183094211368502961688615079$} & $31$ \\ 
$28$ & \multicolumn{1}{l}{$183.8799743836545477274854595901$} & $30$ \\ 
$30$ & \multicolumn{1}{l}{$201.591594898617280704364887093$} & $29$ \\ 
$32$ & \multicolumn{1}{l}{$219.70765025504812387270954474$} & $28$ \\ 
$34$ & \multicolumn{1}{l}{$238.211099545239913239025204$} & $26$ \\ 
$36$ & \multicolumn{1}{l}{$257.0865867898723853694145$} & $24$ \\ 
$38$ & \multicolumn{1}{l}{$276.320189463316984578025$} & $23$ \\ 
$40$ & \multicolumn{1}{l}{$295.8992158830795478713$} & $21$ \\ \hline\hline
\end{tabular}%
\caption{Estimates of even excited state energies fo $\protect\beta =-1,$ $%
m=2$ as obtained by the mapping method with $M=100$. For each $n$ a second
line displays the continuation of digits. The last column displays an
approximate value of the number of significant figures obtained in each
case. }
\label{Tab7}
\end{table}

\begin{table}[tbp]
\centering
\par
\begin{tabular}{ccc}
\hline\hline
$n$ & $E_{n}$ & $N_{c}$ \\ \hline
$1$ & \multicolumn{1}{l}{$%
2.8345362021193042146546762087489649582169401534534235$} & $53$ \\ 
$3$ & \multicolumn{1}{l}{$%
10.0386461207115760433569011502950525317232673587039124$} & $54$ \\ 
$5$ & \multicolumn{1}{l}{$%
19.085714685024187731696964367571874292884831654835817$} & $53$ \\ 
$7$ & \multicolumn{1}{l}{$%
29.462855914201380473167848993120172078754688521382 $} & $50$ \\ 
$9$ & \multicolumn{1}{l}{$%
40.903856271824743299465535879294203331299766065057 $} & $50$ \\ 
$11$ & \multicolumn{1}{l}{$53.245474927626236240667743322663042009574303227$}
& $47$ \\ 
$13$ & \multicolumn{1}{l}{$66.37549586870572508559449815950258220440202765$}
& $46$ \\ 
$15$ & \multicolumn{1}{l}{$80.2111714281336647731118554099129755086234044$}
& $45$ \\ 
$17$ & \multicolumn{1}{l}{$94.68850493291050353683724156050263891174769$} & $%
43$ \\ 
$19$ & \multicolumn{1}{l}{$109.7562587882040753137041026729543093207987$} & $%
43$ \\ 
$21$ & \multicolumn{1}{l}{$125.37231685063352548649262185881882511079$} & $%
41 $ \\ 
$23$ & \multicolumn{1}{l}{$141.501336342288479447357557169786629457$} & $39$
\\ 
$25$ & \multicolumn{1}{l}{$158.11315824690269974081471866579903203$} & $38$
\\ 
$27$ & \multicolumn{1}{l}{$175.181689387733502622384177110426293$} & $36$ \\ 
$29$ & \multicolumn{1}{l}{$192.68409129285960921805894643940927$} & $35$ \\ 
$31$ & \multicolumn{1}{l}{$210.6001761125520587469562806352033$} & $34$ \\ 
$33$ & \multicolumn{1}{l}{$228.91194667874624570198226103433$} & $32$ \\ 
$35$ & \multicolumn{1}{l}{$247.6032395967127362377028468166$} & $31$ \\ 
$37$ & \multicolumn{1}{l}{$266.65944368040870990077882183$} & $29$ \\ 
$39$ & \multicolumn{1}{l}{$286.0672745885394971391857558$} & $28$ \\ 
$41$ & \multicolumn{1}{l}{$305.81459212049335719581218$} & $26$ \\ 
\hline\hline
\end{tabular}%
\caption{Estimates of odd excited state energies fo $\protect\beta =-1,$ $%
m=2 $ as obtained by the mapping method with $M=100$. For each $n$ a second
line displays the continuation of digits. The last column displays an
approximate value of the number of significant figures obtained in each
case. }
\label{Tab8}
\end{table}

These results are in agreement with the estimates given in the literature 
\cite{6395,6342,6431,6432,6339} at least up to the significant figures they
quote.


\begin{thebibliography}{99}
\bibitem{computer} The calculations are performed, with the help of
Mathematica 5.0.1.0. running on a laptop with a processor Intel Pentium M
2.00 GHz, 1.00 Go RAM
\end{thebibliography}
\end{document}